\documentclass[12pt]{article}
\usepackage{graphicx}
\usepackage{amsmath}
\usepackage{amsmath}
\usepackage{latexsym}
\usepackage{bbm}
\usepackage{amssymb}
\usepackage{lipsum}
\usepackage{mathtools}
\usepackage{mathdots}

\usepackage{tikz}
\usetikzlibrary{decorations.pathmorphing}
\usetikzlibrary{decorations.markings}
\usepackage{tikz}
\usetikzlibrary{arrows,shapes}
\usetikzlibrary{trees}
\usetikzlibrary{patterns}
\usetikzlibrary{matrix,arrows} 				
\usetikzlibrary{positioning}				
\usetikzlibrary{calc,through}				
\usetikzlibrary{decorations.pathreplacing}  
\usepackage{pgffor}							

\usetikzlibrary{decorations.pathmorphing}	
\usetikzlibrary{decorations.markings}
\tikzset{
    4Dline/.style={decorate, decoration={snake}, draw},
	provector/.style={decorate, decoration={snake,amplitude=2.5pt}, draw},
        smallvector/.style={decorate, decoration={snake,amplitude=1.5pt,post length=0.5mm}, draw},
    6Dline/.style={draw=black, postaction={decorate},
        decoration={markings,mark=at position .55 with {\arrow[draw=black]{>}}}},
    6Dlinenoarrow/.style={draw=black},
    scalar/.style={dashed,draw=black, postaction={decorate},
        decoration={markings,mark=at position .55 with {\arrow[draw=black]{>}}}},
}

\tikzstyle{block} = [draw, rectangle, 
    minimum height=3em, minimum width=6earticlem]

\usepackage{hyperref}
\usepackage[all]{hypcap}
\usepackage{cite}

\def\tell{\tilde{\ell}}

\def \slc {\text{SL}(2, \mathbb{C})}
\def\a{\alpha}

\def\d{\delta}

\def\e{\varepsilon}

\def\l{\lambda}

\def\s{\sigma}

\def\pa{\partial}

\def\and{{\rm and}}

\newcommand{\be}{\begin{equation}}
\newcommand{\bea}{\begin{eqnarray}}

\newcommand{\ee}{\end{equation}}
\newcommand{\eea}{\end{eqnarray}}

\begin{document}
\vspace*{-1.0in}
\thispagestyle{empty}
\begin{flushright}
QMUL-PH-20-07
\end{flushright}

\normalsize
\baselineskip = 18pt
\parskip = 6pt

\vspace{1.0in}

{\Large \begin{center}
{\bf D3-Brane Loop Amplitudes from \\ M5-Brane Tree Amplitudes}
\end{center}}

\vskip 0.2in

\begin{center}
{\large Congkao Wen, and Shun-Qing Zhang}

\vskip 0.15in

\textit{\small Centre for Research in String Theory, School of Physics and Astronomy\\
Queen Mary University of London,
Mile End Road, London, E1 4NS,
United Kingdom}\\

\end{center}
\vspace{.25in}

\begin{center}
\textbf{Abstract}
\end{center}
\begin{quotation}

We study loop corrections to scattering amplitudes in the world-volume theory of a probe D3-brane, which is described by the supersymmetric Dirac-Born-Infeld theory. We show that the D3-brane loop superamplitudes can be obtained from the tree-level superamplitudes in the world-volume theory of a probe M5-brane (or D5-brane). The M5-brane theory describes self-interactions of an abelian tensor supermultiplet with $(2,0)$ supersymmetry, and the tree-level superamplitudes are given by a twistor formula. We apply the construction to the maximally-helicity-violating (MHV) amplitudes in the D3-brane theory at one-loop order, which are purely rational terms (except for the four-point amplitude). The results are further confirmed by generalised unitarity methods.  Through a supersymmetry reduction on the M5-brane tree-level superamplitudes, we also construct one-loop corrections to the non-supersymmetric D3-brane amplitudes, which agree with the known results in the literature.  

\end{quotation}

\vspace{1cm}

\thispagestyle{empty}

\vfill
\vskip 5.mm
\hrule width 5.cm
\vskip 2.mm
{
\noindent  {\scriptsize e-mails:  {\tt c.wen@qmul.ac.uk, shun-qing.zhang@qmul.ac.uk} }
}

\setcounter{footnote}{0}
\setcounter{page}{0}

\newpage

\tableofcontents

\newpage

\section{Introduction}

The world-volume effective theories of probe branes play important roles in superstring theory and M-theory. This paper studies loop corrections to scattering amplitudes in the world-volume theory of a probe D3-brane of type IIB superstring theory in a 10D Minkowski-space background. The effective field theory is described by the Dirac-Born-Infeld theory with 4D $\mathcal{N}=4$ supersymmetry \cite{Aganagic:1996nn, Cederwall:1996pv, Aganagic:1996pe, Cederwall:1996ri, Bergshoeff:1996tu, Bergshoeff:2013pia}, and we will refer it as the D3-brane theory.  We will argue that the loop corrections to the scattering amplitudes in the D3-brane theory can be obtained from the tree-level amplitudes in theories in higher dimensions. In particular, the relevant higher-dimensional theories are the world-volume theories of a probe M5-brane (of M-theory) in an 11D Minkowski-space background or a probe D5-brane (of type IIB superstring theory) in a 10D Minkowski-space background. We will refer to these two 6D theories as the M5-brane theory and the D5-brane theory, respectively.

The tree-level superamplitudes in the M5-brane theory and D5-brane theory are described by twistor formulations either based on rational maps~\cite{Heydeman:2017yww, Cachazo:2018hqa, Heydeman:2018dje}, or based on the polarised scattering equations~\cite{Geyer:2018xgb, Albonico:2020mge} \footnote{The 6D polarised scattering equations have been extended to scattering amplitudes in 10D and 11D supersymmetric
theories~\cite{Geyer:2019ayz}.} (see also \cite{Schwarz:2020emu} for a recent review on the M5-brane theory and the twistor formulations for the tree-level amplitudes in the theory). These different forms of twistor formulations were later unified in the picture of the Symplectic Grassmannian representation of the 6D superamplitudes~\cite{Schwarz:2019aat}.  The 6D formulae extend the well-known twistor formulation of the scattering amplitudes for 4D $\mathcal{N}=4$
super Yang--Mills (SYM) \cite{Witten:2003nn, Roiban:2004yf}. The twistor formulations have been applied to the tree-level superamplitudes in a variety of supersymmetric theories in 6D. Besides the M5-brane theory and D5-brane theory that we have been discussing, it also describes 6D maximal super Yang-Mills theory, as well as the supergravity theories with $(2,2)$ and $(2,0)$ supersymmetries~\cite{Cachazo:2018hqa, Heydeman:2018dje}.

The D3-brane theory, M5-brane theory, and the D5-brane theory are known to be closely related, as predicted by superstring theory and M-theory. Specifically, both of the 6D theories (the D5 and M5-brane theory) can be truncated (by a procedure of dimensional reduction) to give rise to the 4D theory (the D3-brane theory). The truncation reduces the M5-brane (or D5-brane) tree-level amplitudes in 6D to the D3-brane tree-level amplitudes in 4D~\cite{Heydeman:2017yww}. The tree-level scattering amplitudes in the D3-brane theory have many interesting properties, just to name a few here: the only non-trivial tree-level amplitudes in the D3-brane theory are those with the same number of minus-helicity and plus-helicity photons, namely the helicity is conserved~\cite{Rosly:2002jt}; the amplitudes obey soft theorems, which are very strong constraints that allow to uniquely fix all the tree-level amplitudes in the theory~\cite{Cheung:2014dqa, Cheung:2015ota, Luo:2015tat, Cheung:2016drk, Cheung:2018oki, Rodina:2018pcb}; furthermore, the D-brane tree-level amplitudes are an important part of the so-called unifying relations that relate tree amplitudes in a wide range of effective field theories \cite{Cheung:2017ems}.

It is not surprising that, when dimensionally reduced to 4D, these twistor formulations for 6D tree-level amplitudes reproduce the corresponding formulae for the 4D tree-level amplitudes \cite{He:2016vfi}. This paper considers constructing loop corrections to the scattering amplitudes in the 4D theories from the tree-level amplitudes in 6D theories. In particular, we will construct loop corrections to the amplitudes in the D3-brane theory from the M5-brane (or D5-brane) tree-level amplitudes. As a warm up example, we will also consider loop corrections to the scattering amplitudes in 4D $\mathcal{N}=4$ SYM and $\mathcal{N}=8$ supergravity from the tree-level amplitudes in 6D SYM and supergravity, respectively. In the case of 4D $\mathcal{N}=4$ SYM and $\mathcal{N}=8$ supergravity, the construction reproduces known results in the literature.  

With the help of CHY formulation \cite{Cachazo:2013hca, Cachazo:2013iea} of scattering amplitudes, and built on earlier works from ambi-twistor string theory \cite{Mason:2013sva, Adamo:2013tsa}, it is known that the loop corrections in lower dimensions can be obtained from those in higher dimensions via a peculiar dimensional reduction \cite{Geyer:2015bja, Geyer:2015jch, He:2015yua, Cachazo:2015aol} (see also \cite{CaronHuot:2010zt}).  In particular, for constructing a $n$-point one-loop amplitudes in lower dimensions, we begin with a $(n{+}2)$-point tree-level amplitudes in the corresponding theory in the higher dimensions, and set $n$ of the external momenta in the lower dimensions whereas the remaining two momenta are taken to be forward and stay in the higher dimensions. The forward momenta play the role of the loop momenta of the one-loop amplitudes in the theory in the lower dimensions. Analogous constructions for the two-loop amplitudes have been pushed forward in \cite{Geyer:2016wjx, Geyer:2018xwu, Geyer:2019hnn}, where two pairs of forward momenta become the loop momenta of the two-loop amplitudes.  Instead of using the general dimensional CHY formulations, we will apply the 6D twistor formulations in~\cite{Heydeman:2017yww, Cachazo:2018hqa, Heydeman:2018dje, Geyer:2018xgb, Albonico:2020mge, Schwarz:2019aat} to construct loop corrections to amplitudes in 4D. The advantage of using the 6D twistor formulae is that they make the supersymmetry manifest, and allow us to conveniently utilise the spinor-helicity formalism, which is a powerful tool for computing scattering amplitudes in 4D theories (see, for instance, \cite{Dixon:1996wi, Elvang:2013cua}, for a review.)

The rest of the paper is organised as follows. Section~\ref{sec:6D-SE} reviews the 6D scattering equations, and their applications to the tree-level superamplitudes in various 6D supersymmetric theories.  Section~\ref{sec:forward} discusses the construction of loop superamplitudes in 4D theories from the twistor formulae of 6D tree-level superamplitudes with appropriate forward limits. To illustrate the ideas, we take the one- and two-loop four-point amplitudes in $\mathcal{N}=4$ SYM and $\mathcal{N}=8$ supergravity as examples.  In section~\ref{sec:D3loop}, we apply the ideas to construct loop corrections to the D3-brane theory. In particular, we focus on the maximally-helicity-violating (MHV) amplitudes at one loop, which are given by contact rational terms. The same results are achieved using generalised unitarity methods. We compute explicitly the four- and six-point amplitudes, and also comment on the structure of the $n$-point MHV amplitude. In section~\ref{sec:nonsusy}, through a supersymmetric reduction of 6D tree-level superamplitudes in the M5-brane theory, we study one-loop corrections to the non-supersymmetric D3-brane amplitudes. We further show that our results are in agreement with those in the reference~\cite{Elvang:2019twd}, which were obtained recently by Elvang, Hadjiantonis, Jones and Paranjape using generalised unitarity methods. 

\section{6D twistor formulations} 
\label{sec:6D-SE}

In this section, we will briefly review the twistor formulations for tree-level superamplitudes in 6D theories following the references~\cite{Heydeman:2017yww, Cachazo:2018hqa, Heydeman:2018dje, Geyer:2018xgb, Schwarz:2019aat}.  These twistor formulae of 6D tree-level amplitudes will provide the basis for constructing loop corrections to the amplitudes in 4D theories using forward limits. 

\subsection{6D scattering equations}

The scattering amplitudes of massless particles in 6D are described in terms of the 6D spinor-helicity formalism~\cite{Cheung:2009dc}, which expresses the 6D massless momentum as
\begin{align} \label{eq:6Dspinor-helicity}
p^{AB} = \langle \lambda^A \lambda^{B} \rangle = \frac{1}{2} \epsilon^{ABCD} [ \tilde{\lambda}_{C}
\tilde{\lambda}_{D} ] \, ,
\end{align}
where $A, B=1,2,3,4$ are spinor indices of Lorentz group ${\rm Spin}(1,5)$. Here we have used the short-hand notations 
\begin{align}
 \langle \lambda^A \lambda^{B} \rangle := \lambda^A_a \lambda^{B}_b \epsilon^{ab} \, , \quad
 [ \tilde{\lambda}_{A} \tilde{\lambda}_{B} ]: =\tilde{\lambda}_{A, \hat{a} } \tilde{\lambda}_{B, \hat{b}}
 \epsilon^{\hat{a} \hat{b} } \, ,
\end{align}
where $a, b$ and  $\hat{a}, \hat{b}$ are little-group indices. For a massless particle in 6D, the little group
is ${\rm Spin}(4)\sim {\rm SU}(2)_L \times {\rm SU}(2)_R$, so $a, b$ in the above equation are the indices
of ${\rm SU}(2)_L$ with $a, b=1, 2$,  and  $\hat{a}, \hat{b}=\hat{1}, \hat{2}$ refer to ${\rm SU}(2)_R$.  

The kinematics of massless particles in 6D can be nicely described by the 6D scattering equations, either via rational maps \cite{Heydeman:2017yww, Cachazo:2018hqa, Heydeman:2018dje} or equivalently the polarised scattering equations \cite{Geyer:2018xgb}. We will utilise the polarised scattering equations in our discussions,\footnote{All the discussions also apply to the formalism based on rational maps, see the reference \cite{Schwarz:2019aat} for a detailed discussion of rational maps and their equivalence to the polarised scattering equations via a symplectic Grassmannian.}  which take the following form
\begin{align} \label{eq:polarized-eq}
\int d \mu_n^{6D} = \int  {\prod_{i=1}^n d \s_i \, d^{2} v_i \, d^{2} u_i \over {\rm vol} ( \slc_\s
\times  \slc_u  ) }\prod_{i=1}^n \delta \left( \langle v_{i} \, \epsilon_i \rangle -1 \right) \, \delta^4 \left( \langle v_{i}
\lambda^{A}_i \rangle - \langle u_{i}\lambda^{A} (\s_i) \rangle  \right) \, ,
\end{align}
where we mod out the $ \slc$ symmetries acting on world-sheet coordinates $\s_i$ as well as on the coordinates $u_i$. 
The rational functions $\lambda^{A a} (\s)$ are given by
\begin{align}
\lambda^{A a} (\s)  = \sum_{j=1}^n { u^a_{j} \langle \epsilon_{j} \l^{A}_j \rangle \over \s - \s_j }\, .
\end{align}
Here $\l^{Ab}_j$ are the 6D helicity spinors that we introduced in (\ref{eq:6Dspinor-helicity}), and the constraints in (\ref{eq:polarized-eq}) implies the momentum conservation, i.e. $\sum_{i=1}^n \langle \l^{A}_i \l^{B}_i \rangle =0$. 
It is convenient to choose the little-group spinors $\epsilon_i$ that enter in the constraints $\langle v_{i} \, \epsilon_i \rangle =1$ to be $\epsilon_{i, a}= (0, 1)$. For such a choice, the delta-function constraints $\langle v_{i} \, \epsilon_i \rangle =1$ are solved by $v_{i, a}= (1, v_i)$.    

It is worth noticing that (\ref{eq:polarized-eq}) can be recast into a matrix form, 
\bea \label{eq:Symplectic-V}
\int d \mu_n^{6D} = \int  {\prod_{i=1}^n d \s_i \, d^{2} v_i \, d^{2} u_i
\over {\rm vol} ( \slc_\s \times  \slc_u  ) }\prod_{i=1}^n
\delta( \langle v_{i} \, \epsilon_i \rangle -1)\delta^{4} ( V \cdot \Omega \cdot  \Lambda^A) \, ,
\eea
where $\Lambda^A$ is a $2n$-dimensional vector encoding the external helicity spinors, 
\begin{equation} \label{eq:Lambda}
\Lambda^A :=\{ \lambda^{A}_{1, 1}, \lambda^{A}_{2, 1}, \ldots, \lambda^{A}_{n, 1},  \lambda^{A}_{1, 2},
\lambda^{A}_{2, 2}, \ldots, \lambda^{A}_{n, 2}\}  \,  .
\end{equation}
and $V$ is a $n\times 2n$ matrix  that follows from (\ref{eq:polarized-eq})
\bea \label{eq:V-matrix}
V_{i; j, a}= \left\{
                \begin{array}{ll}
                  v_{i, a} \qquad \qquad \quad {\rm if} \qquad  i=j \\
       - { \langle u_i u_j \rangle \over \s_{ij} } \epsilon_{j, a}   \qquad   {\rm if} \qquad i \neq j ,
                \end{array}
              \right.
\eea
with $\s_{ij}:= \s_{i} - \s_{j}$, and $\Omega$ is the symplectic metric
\begin{equation}
\Omega= \begin{pmatrix}
0 & \mathbb{I}_n \\
- \mathbb{I}_n  & 0
\end{pmatrix} \, . 
\end{equation}
Importantly, under the condition $\langle v_{i} \, \epsilon_i \rangle =1$,  the matrix $V$ obeys the following symplectic condition~\cite{Schwarz:2019aat}, 
\bea
V \cdot \Omega \cdot V^T =0 \, . 
\eea
Therefore the space of the matrices $V$ forms the symplectic Grassmannian~\cite{Karpman:2015oha, stratificationLG}.  Finally, one may define a different set of 6D scattering equations with the helicity spinors $\tilde{\lambda}_{i, A}^{ \hat a}$, 
\bea \label{eq:Symplectic-tV}
\int d \tilde{\mu}_n^{6D} = \int  {\prod_{i=1}^n d \s_i \, d^{2} \tilde{v}_i \, d^{2} \tilde{u}_i
\over {\rm vol} ( \slc_\s \times  \slc_u  ) }\prod_{i=1}^n
\delta( [ \tilde{v}_{i} \, \tilde{\epsilon}_i ] -1)\delta^{4} ( \tilde{V} \cdot \Omega \cdot  \tilde{\Lambda}_A) \, ,
\eea
which is needed for determining $\tilde{v}_i, \tilde{u}_i$. The variables $\tilde{v}_i, \tilde{u}_i$ are required for constructing superamplitudes of the non-chiral theories such as 6D SYM, D5-brane theory and supergravity theories. 

\subsection{6D tree-level superamplitudes}

To describe the on-shell supersymmetry in 6D, we further introduce Grassmann variables, $\eta^I_{i, a}$ and $\tilde{\eta}^{\tilde I}_{i, \hat a}$, with $I=1, 2, \cdots, \mathcal{N}$ and $\tilde{I}=1, 2, \cdots, \widetilde{\mathcal{N}}$ for a theory with $(\mathcal{N}, \widetilde{\mathcal{N}})$ supersymmetry, and $i$ is the particle label. In particular, the supercharges are defined by
\begin{align}
q^{A, I}_i= \langle \lambda^A_i \, \eta^I_i \rangle\, , \qquad \bar{q}^A_{I, i} = \lambda^A_{i, a} {\partial \over \partial \eta^I_{i, a} }\, , 
\end{align}
and similarly 
\begin{align}
\tilde{q}_{A, i}^{\tilde I}= [\tilde{\lambda}_{A, i}\, \tilde{\eta}^{\tilde I}_i ]\, , \qquad \bar{\tilde{q}}_{A, i, \tilde I} = \tilde{\lambda}_{i, A, \hat a} {\partial \over \partial  \tilde{\eta}^{\tilde I}_{i, \hat a} } \, . 
\end{align}
We note the supercharges obey the correct supersymmetry algebra, $\{q^{A, I}_i, \bar{q}^B_{J, i}\} =\delta^I_J \, p^{AB}_{i}$, and similar algebra relations for $\tilde{q}_{A, i}^{\tilde I}, \bar{\tilde{q}}_{A, i, \tilde I}$. The superamplitudes should be annihilated by the total supercharges, namely $Q^{A, I}_n = \sum_{i=1}^n q^{A, I}_i,\, \bar{Q}^{A, I}_n = \sum_{i=1}^n \bar{q}^{A, I}_i$. Therefore, the $n$-point superamplitude must be proportional to $\delta^{4 \mathcal{N} }(Q_n ) \, \delta^{4 \widetilde{\mathcal{N} } }(\bar{Q}_n ) \, .$

The on-shell spectrum of a supersymmetric theory can be packaged into an on-shell superfield with the help of the Grassmann variables, $\eta^I_{i, a}$ and $\tilde{\eta}^{\tilde I}_{i, \hat a}$, and the on-shell superamplitudes are functions of  helicity spinors as well as the Grassmann variables. Let us begin with the 6D SYM with $(1,1)$ supersymmetry. The on-shell spectrum of the theory is given by 
\bea \label{eq:6DSYM}
\Phi(\eta, \bar{\eta})= \phi^{1\hat{1}} + \eta_a \psi^{a \hat{1}}+ \tilde{\eta}_{\hat a} \psi^{1 \hat {a} }
+ \eta_a \tilde{\eta}_{\hat a}  A^{a \hat{a}} + \ldots
+ ( \eta)^2(\tilde{\eta})^2 \phi^{2\hat{2}} \, ,
\eea
where, for instance, $ \phi^{1\hat{1}}$ is one of four scalars of the theory, and $A^{a \hat{a}}$ is the 6D gluon. Similar to the CHY construction of scattering amplitudes \cite{Cachazo:2013hca, Cachazo:2013iea}, the tree-level amplitudes of a generic 6D theory take the following form in the twistor formulations
\begin{align}
\mathcal{A}_n = \int d\mu^{\rm 6D}_n \, {\cal I}_L   \, {\cal I}_R \, .
\end{align}
In the above formula, the measure $d\mu^{\rm 6D}_n$ imposes the 6D scattering equations which are given in (\ref{eq:polarized-eq}) or (\ref{eq:Symplectic-V}), and ${\cal I}_L$ and $ {\cal I}_R$ specify the dynamics of the theory, which will be called as left and right integrand. In the case of 6D $(1,1)$ SYM, the  twistor formula of the tree-level superamplitudes is given as
\begin{align} \label{eq:even-YM}
\mathcal{A}^{\rm SYM}_{n}  (\alpha)
=
\int d\mu^{\rm 6D}_n \, {\cal I}_L^{(1,1)}   \, {\cal I}_R^{(\alpha)} \, ,
\end{align}
where the left and right integrands take the following form
\bea
{\cal I}_L^{(1,1)} = \delta^n (V \cdot \Omega \cdot \eta) \delta^n (\widetilde{V} \cdot \widetilde{\Omega} \cdot
\tilde{\eta})\,  {\rm det}^{\prime} H_n \, , \qquad {\cal I}_R^{(\alpha)} = {\rm PT} (\alpha) \, .
\eea
Here $\eta, \tilde{\eta}$ are the Grassmann version of $\Lambda^A$ in (\ref{eq:Lambda}), 
\bea
 \eta &=& \{ \eta_{1, 1}, \eta_{2, 1}, \ldots, \eta_{n, 1},  \eta_{1, 2}, \eta_{2, 2}, \ldots, \eta_{n, 2}\}\, , \cr
\tilde{\eta} &=& \{\tilde{\eta}_{1, \hat 1}, \tilde{\eta}_{2, \hat 1}, \ldots, \tilde{\eta}_{n, \hat 1}, \tilde{\eta}_{1, \hat 2}, \tilde{\eta}_{2, \hat 2}, \ldots, \tilde{\eta}_{n,  \hat 2}\} \, ,
\eea
and the Grassmann delta functions in ${\cal I}_L^{(1,1)} $ imply the conservation of supercharges of 6D $(1,1)$ supersymmetry. 

Other building blocks of the integrands are defined as below. First, ${\rm PT} (\alpha)$ is the Parke--Taylor factor, which encodes the colour structure of Yang--Mills amplitudes. The notation $\alpha$ represents a permutation of the external particles $\{1,2, \ldots, n\}$. For instance, when $\alpha$ is the identity permutation,
\bea
{\rm PT} (1,2, \ldots, n) = {1\over \sigma_{12} \sigma_{23} \cdots \sigma_{n-1\, n} \sigma_{n1}} \, .
\eea
The $n\times n$ matrix $H_n$ has the following entries \cite{Geyer:2018xgb}
\bea
H_{ij} = { \langle \epsilon_i  \lambda^A_i \rangle [ \tilde{\epsilon}_j  \tilde{\lambda}_{A,j}  ] \over
\sigma_{ij} } \quad {\rm for}\quad  i \neq j \, , \qquad
u_{i, a} H_{ii} = - {{\lambda}^A_{a}(\sigma_i) [ \tilde{\epsilon}_i  \tilde{\lambda}_{A,i}  ]  } \, .
\eea
Here, just as $\epsilon_{i,a}$, we can choose $\tilde{\epsilon}_{i, \hat{a}} = (0, 1)$. Note that $H_{ii}$ is
independent of the choice of little-group index $a$, namely it is a Lorentz scalar. The reduced determinant
${\rm det}^{\prime}H$ in ${\cal I}_L^{(1,1)}$ is defined as
\bea
{\rm det}^{\prime} H = { {\rm det} H^{[ij]}_{[kl]}  \over \langle u_i u_j\rangle [\tilde{u}_k \tilde{u}_l]}
\, ,
\eea
where $H^{[ij]}_{[kl]}$ means that we remove the $i$-th and $j$-th columns as well as the $k$-th and $l$-th
rows, and the result is independent of the choices of $i,j$ and $k,l$.  Note that the conjugate variables such as $\tilde{v}_i, \tilde{u}_i$ appeared in the integrands are determined by scattering equations in (\ref{eq:Symplectic-tV}). 

The tree-level superamplitudes in 6D supergravity with $(2,2)$ supersymmetry are obtained by the double copy of the $(1,1)$ SYM superamplitudes. We have, 
\begin{align} \label{eq:22SUGRA}
\mathcal{M}^{(2,2)}_{n}
&=\int d\mu_n^{\rm 6D}  \delta^{2 \times n} (V \cdot \Omega \cdot \eta^I) \delta^{2 \times n}
(\widetilde{V} \cdot  \widetilde{\Omega}  \cdot \tilde{\eta}^{\tilde I})  \,
({\rm det}^{\prime} H_n)^2 \, ,
\end{align}
where $I=1,2$ and $\tilde{I}=\tilde{1}, \tilde{2}$ for the 6D $(2,2)$ supersymmetry. 

Let us turn to the M5-brane theory and D5-brane theory. The D5-brane theory has $(1,1)$ supersymmetry and contains the same spectrum as SYM, that is given in (\ref{eq:6DSYM}). The M5-brane theory is a chiral theory with $(2,0)$ supersymmetry, and the on-shell superfield is a tensor multiplet, which is given as
\bea \label{eq:M5-field}
\Phi(\eta) = \phi + \eta^a_I \psi_a^I + \eta^a_I  \eta^{I, b}  B_{ab} + \ldots + (\eta)^4 \bar{\phi} \, ,
\eea
with $I=1,2$, and $B_{ab}$ is the on-shell 6D self-dual tensor. For the D5-brane theory and the M5-brane theory, only the even-multiplicity amplitudes are non-trivial, and the odd-multiplicity ones vanish identically. The left integrands for the M5-brane theory and the D5-brane theory of the twistor formulations are in fact the same,
\begin{equation} \label{M5-L}
\mathcal{I}_L^{\text{M5}}=\mathcal{I}_L^{\text{D5}}=(\text{Pf}' S_n)^2 \, ,
\end{equation}
where $S_n$, which  is only defined for even $n$, is an $n \times n$ matrix with entries given as
\bea
[S_n]_{ij} = {p_i \cdot p_j \over \s_{ij}} \, ,
\eea
with $p_i, p_j$ being the 6D momenta. The reduced Pfaffian of $S_n$, $\text{Pf}' S_n$,  is defined as
\bea  
{\rm Pf}^{\prime} S_n = {(-1)^{k+l} \over \sigma_{kl}} {\rm Pf} (S_n)^{kl}_{kl} \, ,
\eea
where $(S_n)^{kl}_{kl}$ is an $(n{-}2) \times (n{-}2)$ matrix with the $k$-th and $l$-th rows and columns of $S_n$ removed, and the result is independent of the choice of $k, l$. The right integrand of the D5-brane superamplitudes is given by
\begin{align} \label{D5-R}
\mathcal{I}_R^{\text{D5}}&=  \delta^n \left( V \cdot \Omega \cdot \eta \right)\delta^n \left( \tilde{V} \cdot \Omega \cdot \tilde{\eta} \right) (\text{Pf} \, U_n)^{-\frac{1}{2}}(\text{Pf}\, \tilde{U}_n)^{-\frac{1}{2}} \, \text{Pf}' \, S_n\, ,
\end{align}
whereas the right integrand of the M5-brane superamplitudes takes a simpler form,
\begin{align} \label{M5-R}
\mathcal{I}_R^{\text{M5}}&= \delta^{2 \times n} \left( V \cdot \Omega \cdot \eta^I \right) ({\text{Pf}\, U_n})^{-1} \,  \text{Pf}' \, S_n  \, .
\end{align}
The fermionic delta functions in $\mathcal{I}_R^{\text{D5}}$ lead to the $(1,1)$ supersymmetry for the D5-brane theory, and the fermionic delta functions in $\mathcal{I}_R^{\text{M5}}$ encode $(2,0)$ supersymmetry for the M5-brane theory. The object $U_{n}$ that appears in the above formulae is an $n \times n$ anti-symmetric matrix with entries given by
\be
U_{ij} =\frac{\langle u_i \, u_j \rangle^2}{\s_{i j}} \, ,
\ee
and one may define the conjugate matrix $\tilde{U}_n$ in a similar fashion using $\tilde{u}_i$. 
Note that $\text{Pf}' \, S_n$ is related to $\det' H_n$ through the following identity
\bea
(\text{Pf} \, U_n)^{-\frac{1}{2}}(\text{Pf}\, \tilde{U}_n)^{-\frac{1}{2}} \, \text{Pf}' \, S_n  = \text{det}' H_n \, ,
\eea
which is true under the support of 6D scattering equations.  
In summary, the tree-level superamplitudes in the M5-brane theory is given by
\bea \label{eq:M5}
\mathcal{A}^{\rm M5}_n = \int d \mu^{\rm 6D}_n \, \mathcal{I}_L^{\text{M5}} \, \mathcal{I}_R^{\text{M5}} \, ,
\eea
with $ \mathcal{I}_L^{\text{M5}}$ and $\mathcal{I}_R^{\text{M5}}$ given in (\ref{M5-L}) and (\ref{M5-R}), respectively. A similar formula can be written down for the D5-brane superamplitudes. From the twistor formulae, we can obtain explicit superamplitudes, for instance, the four-point amplitude of M5-brane theory is simply
\bea \label{eq:A4}
\mathcal{A}_4^{\rm M5} = \delta^{8} (Q_4) \, , 
\eea
where recalling that $Q_4$ is the supercharge, and it is defined as $Q^{A, I}_4 =\sum_{i=1}^4 \langle \lambda_i^A \, \eta^I_i \rangle$. 

\subsection{D3-brane massive tree-level amplitudes}
\label{sec:tree-amplitude}

After a dimensional reduction to 4D, the M5-brane superfield defined in (\ref{eq:M5-field}) reduces to the D3-brane superfield, which is identical to the superfield of $\mathcal{N}=4$ SYM, given as (in the non-chiral form),  
\begin{align}  \label{eq:non-chiral_form}
\Phi_{\mathcal{N}=4}(\eta_+, {\eta}_-) = \phi + \eta^{\tilde{I}}_+ \psi^+_{\tilde I} +\eta^{I}_- \psi^-_{I} + \cdots +(\eta_+)^2 A^+ +(\eta_-)^2 A^-+ \cdots + (\eta_+)^2 (\eta_-)^{2}  \bar{\phi}\, ,
\end{align}
where we have identified $\{ \eta_{1}, \tilde{\eta}_{\hat 1} \}$ of 6D $(1,1)$ supersymmetry  as $\eta^{I}_-$ with $I=1,2$, and $\{\eta_{2}, \tilde{\eta}_{\hat 2} \}$ as $\eta^{\tilde I}_+$ with ${\tilde I}=1,2$. The supercharges are given by
\bea \label{eq:4D-charge}
Q^{\alpha, I}_n= \sum_{i=1}^n \lambda_i^{\alpha} \eta^I_{i, -}\, , \qquad \tilde{Q}^{\dot \alpha, \tilde I}_n= \sum_{i=1}^n \tilde{\lambda}_i^{\dot \alpha}  {\eta}^{\tilde I}_{i, +} \, .
\eea
So $A^{1\hat{1}}$ is identified as the minus-helcity photon (or gluon, in the case of SYM) $A^-$, $A^{2\hat{2}}$ as the plus-helcity photon (or gluon)  $A^+$ etc. 
The tree-level amplitudes in the D3-brane theory are obtained through a dimension reduction by setting all the external momenta of the M5-brane amplitudes in 4D \cite{Heydeman:2017yww}. The dimension reduction procedure leads to the twistor formulations for the tree-level superamplitudes in the D3-brane theory \cite{He:2016vfi}.

Here we are interested in tree-level superamplitudes in the D3-brane theory with massive states, since they are relevant for constructing loop corrections that we will consider in the section \ref{sec:GUM} using generalised unitarity methods. The massive tree-level amplitudes in the D3-brane theory can also be obtained from the  M5-brane tree-level amplitudes by a careful dimension reduction. In particular, the masses may be viewed as extra dimensional momenta as Kaluza-Klein modes. The D3-brane superfield with massive sates is straightforwardly obtained from (\ref{eq:M5-field})  
\bea 
\Phi_{\rm massive}(\eta) = \phi + \eta^a_I \psi_a^I + \eta^a_I  \eta^{I, b} A_{ab} + \ldots + (\eta)^4 \bar{\phi} \, ,
\eea
where $a=1,2$ are the little group indices of massive particles in 4D, for instance $A_{ab}$ is the 4D massive vector. The superamplitudes in the D3-brane theory with both massless and massive states are obtained from (\ref{eq:M5}) by setting the 6D helicity spinors as 
\begin{equation}\label{eq:4dkin}
\lambda_{i,a}^{A}= \left( \begin{array}{cccc}
0  &\lambda^{\alpha}_i \\
\tilde{\lambda}^{\dot{\alpha}}_i  & 0
\end{array}  \right)\, ,
\end{equation}
for the massless states, and
\begin{equation} \label{eq:FD}
\lambda_{j,a}^{A}= \left( \begin{array}{cccc}
\lambda_{j,1}^{\alpha}  &\lambda_{j, 2}^{\alpha}\\
\tilde{\lambda}^{\dot{\alpha}}_{j, 1}  & \tilde{\lambda}^{\dot{\alpha}}_{j, 2}
\end{array}  \right) \, ,
\end{equation}
for the massive states with complexified masses given by  $\mu_j = \langle \lambda_{j}^{1}  \lambda_{j}^{2} \rangle$ and $\tilde{\mu}_j = \langle \tilde{\lambda}^{\dot{1}}_{j}   \tilde{\lambda}^{\dot{2}}_{j} \rangle$.\footnote{The same procedure has been applied to 6D SYM amplitudes to obtain massive amplitudes in 4D $\mathcal{N}=4$ SYM on the Coulomb branch \cite{Cachazo:2018hqa}.} We will mostly consider the case with two massive states (say, they are particles $i$ and $j$), in which case $\mu_i = - \mu_j = \mu$ and $\tilde{\mu}_i= - \tilde{\mu}_j =\tilde{\mu}$, because the extra dimensional momenta should be conserved.   For instance, at four points, from (\ref{eq:A4}), we find the superamplitude with two massive and two massless states is given by
\bea
\mathcal{A}_4 = \delta^4(\lambda^{\a}_{1,a} \eta_1^{I a} + \lambda^{\a}_{2,a} \eta_2^{I a}+
 \lambda^{\a}_{3} \eta_{3,-}^{I} + \lambda^{\a}_{4} \eta_{4,-}^{I} )
 \delta^4(\tilde{\lambda}^{\dot \a}_{1, a} \eta_1^{\tilde I a} + \tilde{\lambda}^{\dot \a}_{2, a} \eta_2^{\tilde I a}+
 \tilde{\lambda}^{\dot \a}_{3} \eta_{3,+}^{\tilde I} + \tilde{\lambda}^{\dot \a}_{4} \eta_{4,+}^{\tilde I} ) \, ,
\eea
where particles $1$ and $2$ are massive, and $3$ and $4$ are massless. From the superamplitude, we can also obtain component amplitudes, for instance, 
\bea
A_4 (\phi_1, \bar{\phi}_2, 3^+, 4^+) = - \mu^2 [3\,4 ]^2\,,
\eea
where $\phi_1, \bar{\phi}_2$ are massive scalars with mass $\mu$.

\begin{figure}
\begin{center}
	{\begin{tikzpicture}[scale=1.2, line width=1 pt]
	
		\draw[4Dline] (0,0)--(-1.4,0);
		\draw[6Dlinenoarrow] (0,0)--(-1,-1);
		\draw[4Dline] (0,0)--(-1,1);
		\draw[4Dline] (2,0)--(3.4,0);
		\draw[6Dlinenoarrow] (2,0)--(3,-1);
		\draw[4Dline] (2,0)--(3,1);
		\draw[6Dlinenoarrow] (0,0)--(2,0);
		\draw[black,fill=lightgray] (0,0) circle (1.8ex);
		\node at (0,0) {$Q_{L}$};
		\draw[black,fill=lightgray] (2,0) circle (1.8ex);
		\node at (2,0) {$Q_{R}$};
		\node at (1, -0.4) {$K$};
		\node at (-1.6,0) {$2$};
		\node at (3.6,0) {$5$};
		\node at (-1.2,-1.2) {$1$};
		\node at (-1.2,1.2) {$3$};
		\node at (3.2,1.2) {$4$};
		\node at (3.2,-1.2) {$6$};
		
		\end{tikzpicture}}
		{\begin{tikzpicture}[scale=1.2, line width=1 pt]	
		\draw[6Dlinenoarrow] (0,0)--(-1.4,0);
		\draw[6Dlinenoarrow] (0,0)--(-1,-1);
		\draw[4Dline] (0,0)--(-1,1);
		\draw[4Dline] (2,0)--(3.4,0);
		\draw[4Dline] (2,0)--(3,-1);
		\draw[4Dline] (2,0)--(3,1);
		\draw[4Dline] (0,0)--(2,0);
		\draw[black,fill=lightgray] (0,0) circle (1.8ex);
		\node at (0,0) {$Q_{L}$};
		\draw[black,fill=lightgray] (2,0) circle (1.8ex);
		\node at (2,0) {$Q_{R}$};
		\node at (1, -0.4) {$K$};
		\node at (-1.6,0) {$1$};
		\node at (3.6,0) {$4$};
		\node at (-1.2,-1.2) {$6$};
		\node at (-1.2,1.2) {$2$};
		\node at (3.2,1.2) {$3$};
		\node at (3.2,-1.2) {$5$};
		
		\end{tikzpicture}}
\caption{The vertices, $Q_{L}$ and $Q_R$, represent four-point superamplitudes $\delta^8(Q_{L})$ and $\delta^8(Q_{R})$. They are glued together by an on-shell propagator $K$. The curvy lines denote massless particles in 4D, while the solid line indicates 4D massive particles. In the above diagram, we choose leg-1 and leg-6 to be massive states in 4D.}
\label{fig:4pt_gluing}
\end{center}
\end{figure}
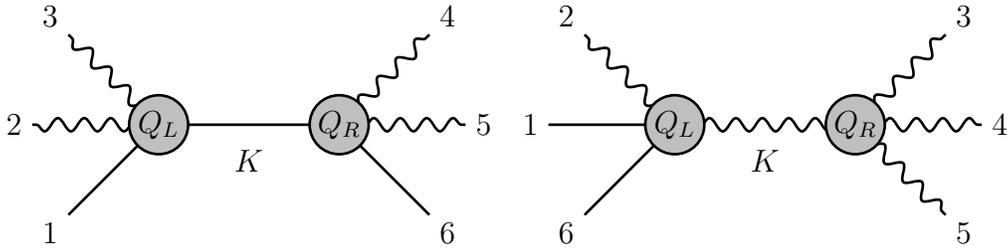

For higher-point amplitudes, we can construct the superamplitude by writing down an ansatz consisting of factorisation terms, which are obtained from lower-point amplitudes, as well as some possible contact terms, and then compare the ansatz with the twistor formula (\ref{eq:M5}) to determine any unfixed parameters. For instance, at six points, the factorisation terms of the tree-level amplitude with two massive states  are shown in Fig.(\ref{fig:4pt_gluing}), where each diagram takes the form of
\bea
\int d^4 \eta_K\, \delta^8(Q_L) {1\over P_{K}^2}  \delta^8(Q_R)  + {\rm Perm} \, . 
\eea
We have used the four-point superamplitude given in (\ref{eq:A4}) with the understanding that kinematics are projected to 4D as described in the above paragraphs. The Grassmann integration is to sum over the intermediate states, and ``${\rm Perm}$" represents summing over all the independent permutations (we will use the same notation in the later sections).  It is straightforward to check that the factorisation terms as shown in Fig.(\ref{fig:4pt_gluing}) agree with (\ref{eq:M5}), it therefore implies that no contact term exists at six points. This is consistent with the known fact that a six-point supersymmetric contact term with six derivatives is not allowed by $(2,0)$ (and $(1,0)$) supersymmetry  \cite{Chen:2015hpa}.  This is also in agreement with what was found in \cite{Elvang:2019twd} for the all-plus and single-minus amplitudes (with two additional massive scalars).\footnote{We should be careful when discussing contact terms because one can always modify the factorisation terms such that the contact terms are changed accordingly. However, here we are talking about superamplitudes and the question whether one can write down a supersymmetric contact term is unambiguous.}

Similar computation applies to higher-point amplitudes. In particular, consider the eight-point superamplitudes, for which the factorisation terms schematically take the form of
\bea \label{eq:8ptASUSY}
\int d^4 \eta_{P_{K_1}}\, d^4 \eta_{P_{K_3}}\, \delta^8(Q_{K_1}) \delta^8(Q_{K_2}) \delta^8(Q_{K_3}) {1\over P_{K_1}^2\, P_{K_3}^2} +{\rm Perm}  \, .
\eea
We find in general contact terms are required to match with the twistor formula  (\ref{eq:M5}), except for the special cases with the helicity configurations being all-plus and single-minus photons (with two additional massive scalars), which agrees with the results of \cite{Elvang:2019twd}.  However, for more general helicity configurations, we do require contact terms. For instance, the amplitude of two-minus photons, with the factorisation terms given in (\ref{eq:8ptASUSY}) (after projecting to the corresponding component amplitude), we find the following contact term is required to match with twistor formula (\ref{eq:M5}), 
\begin{align} \label{eq:A8-tree}
A_{\rm cont}(1^-, 2^-, 3^+, 4^+, 5^+, 6^+, \phi_7, \bar{\phi}_8) &= -6 \mu^2 \langle 12 \rangle^2 ([34]^2[56]^2+ {\rm Perm} ) \\
& = -6 \mu^2 \langle 12 \rangle^2 ([34]^2[56]^2+ [35]^2[46]^2 + [36]^2[45]^2)  \, , \nonumber
\end{align}  
where $\phi_7, \bar{\phi}_8$ are massive scalars with mass $\mu$.  We will come back to this term when we consider its contributions to loop corrections.

\section{Loops from trees} 
\label{sec:forward}

This section will describe the general ingredients for constructing loop amplitudes from tree-level amplitudes in higher dimensions using the twistor formulations. We would like to emphasise that our construction makes the supersymmetry manifest, and utilises the powerful spinor helicity formalism. This is due to the fact that, instead of the general-dimensional CHY formulae, we will apply the 6D twistor formulations that we reviewed in the previous section, from which we will construct the loop corrections to 4D superamplitudes. To illustrate the idea, we will study the well-known one- and two-loop superamplitudes in $\mathcal{N}=4$ SYM and $\mathcal{N}=8$ supergravity using our construction.  The loop corrections to the D3-brane superamplitudes (as well as non-supersymmetric D3-brane amplitudes) will be studied in the following sections.  

\subsection{Loop corrections from higher-dimensional tree amplitudes}

It was argued \cite{He:2015yua, Cachazo:2015aol} that with an appropriate forward limit, the CHY scattering equations for the tree-level amplitudes give rise to the one-loop scattering equations that were originally obtained from the ambi-twistor theory \cite{Mason:2013sva, Adamo:2013tsa}.  To construct the loop corrections to $n$-point scattering amplitudes in lower-dimensional theories,  we start with the $(n{+}2)$-point tree-level amplitudes in corresponding theory in the higher dimensions. We then set $n$ of the external momenta in the lower dimensions of the loop amplitudes that we are interested in, whereas the remaining two momenta, which still stay in the higher dimensions, are taken to be forward. This pair of the forward momenta play the role of the loop momenta of the one-loop amplitudes in the lower dimensions. Analogously, with a careful analysis, the two-loop corrections to $n$-point amplitudes can be obtained from the $(n{+}4)$ tree-level in higher dimensions with two pairs of forward momenta that become the loop momenta of the loop amplitudes \cite{Geyer:2016wjx, Geyer:2018xwu, Geyer:2019hnn}.  

Explicitly, the forward limit procedure in our construction works in the following way: To compute an $n$-point one-loop amplitude in 4D, we set the kinematics (the spinor-helicity variables) of the $(n{+}2)$-point tree-level amplitude in 6D being parametrised as
\begin{equation}\label{eq:4dkin}
\lambda_{i,a}^{A}= \left( \begin{array}{cccc}
0  &\lambda^{\alpha}_i \\
\tilde{\lambda}^{\dot{\alpha}}_i  & 0
\end{array}  \right), \quad \text{for}\;\; i=1,...,n \, ,
\end{equation}
which become the massless external kinematics of the one-loop amplitude in 4D. We also use $k^{\a\,\dot{\a}}_i=\lambda_i^\a\tilde{\lambda}_i^{\dot{\a}}$ to denote massless 4D momenta for external particles. As for the remaining two helicity spinors of the forward-limit particles, we set them to be
\begin{equation} \label{eq:FD}
\lambda_{n+1,a}^{A}= \left( \begin{array}{cccc}
\lambda_1^{\alpha}  &\lambda_2^{\alpha}\\
\tilde{\lambda}^{\dot{\alpha}}_1  & \tilde{\lambda}^{\dot{\alpha}}_2
\end{array}  \right), \quad\lambda_{n+2,a}^{A}= \left( \begin{array}{cccc}
\lambda_1^{\alpha}  &-\lambda_2^{\alpha}\\
\tilde{\lambda}^{\dot{\alpha}}_1  & -\tilde{\lambda}^{\dot{\alpha}}_2
\end{array}  \right)\, ,
\end{equation}
such that $\langle \lambda_{n+1}^{A} \lambda_{n+1}^{B}  \rangle = - \langle \lambda_{n+2}^{A} \lambda_{n+2}^{B}  \rangle$, namely the 6D momenta $p^{AB}_{n+1}, p^{AB}_{n+2}$ are forward and the 4D part is identified as the loop momentum $\ell$ with $\ell^2 \neq 0$.  Furthermore, we should identify the Grassmann variables $\eta_{i}^{I}, \tilde{\eta}_{i}^{\tilde{I}}$ (for $i=1,2, \cdots, n$) with the Grassmann variables of lower-dimensional supersymmetric theories (as we will see explicitly in examples in the following sections). As for the forward-limit pair, we set 
\begin{equation} \label{eq:FW-eta}
\eta_{n+2}^{I, 1}=\eta_{n+1}^{I, 1}\, , \quad \eta_{n+2}^{I, 2}=- \eta_{n+1}^{I, 2}\, , \quad 
\tilde{\eta}_{n+2}^{\tilde{I}, \hat 1}= \tilde{\eta}_{n+1}^{\tilde{I}, \hat 1}, \quad  \tilde{\eta}_{n+2}^ {\tilde{I}, \hat 2}=- \tilde{\eta}_{n+1}^{\tilde{I}, \hat 2} \, ,
\end{equation}
such that the supercharges $q^I_{n+1} = -q^I_{n+2}$ and $\tilde{q}^{\tilde{I}}_{n+1} = -\tilde{q}^{\tilde{I}}_{n+2}$. This is required for the conservation of supercharges of external particles, namely $\sum_{i=1}^n q^I_i = \sum_{i=1}^n \tilde{q}^{\tilde I}_i =0$. 

With this set up, the twistor formulation for the one-loop amplitude is then given as
\begin{align} \label{eq:one-loop-twistor}
\mathcal{A}^{(1)}_{{\rm 4D}, n}
&= \int \frac{d^{4} \ell}{\ell^2} \int d^{2\mathcal{N}}\eta_{n+1}d^{2\widetilde{\mathcal{N}} } \tilde{\eta}_{n+1}\, \mathcal{A}^{(0)}_{{\rm 6D}, n+2} \big{|}_{\rm F.L.} 
\cr
&= \int \frac{d^{4} \ell}{\ell^2} \int   d\mu^{\rm 6D}_{n+2} \, d^{2\mathcal{N}}\eta_{n+1}d^{2 \widetilde{\mathcal{N}}} \tilde{\eta}_{n+1} \, \mathcal{I}_L \mathcal{I}_R \big{|}_{\rm F.L.} \, ,
\end{align}
where we have denoted the $n$-point one-loop amplitude in 4D as $\mathcal{A}^{(1)}_{{\rm 4D}, n}$. The object $\mathcal{A}^{(0)}_{{\rm 6D}, n+2} \big{|}_{\rm F.L.}$ denotes the 6D $(n{+}2)$-point tree-level amplitude with the special forward kinematics given in (\ref{eq:4dkin}) and (\ref{eq:FD}), as well as (\ref{eq:FW-eta}) for the Grassmann variables. The Grassmann integral is to sum over all the internal states for the pair of the particles that are taken to be forward. In the last step of (\ref{eq:one-loop-twistor}), we have expressed the 6D tree-level amplitudes in the twistor formulations, with the measure $d\mu^{\rm 6D}_{n+2}$ given in (\ref{eq:Symplectic-V}), and the precise form of  the integrand $\mathcal{I}_L$ and $\mathcal{I}_R$ depends on the theory we are considering.  As discovered in \cite{Geyer:2015bja} (see also \cite{Baadsgaard:2015twa}),  this formulation typically leads to loop integrands with linear propagators, and we will see more examples of these linear propagators in the following sections.\footnote{For a proposal of obtaining loop integrands with standard quadratic Feynman propagators using CHY formulation, see \cite{Gomez:2017lhy, Gomez:2017cpe, Ahmadiniaz:2018nvr, Agerskov:2019ryp}. }

\subsection{Loop corrections to SYM and supergravity superamplitudes}

To illustrate the ideas, we consider the one- and two-loop amplitudes in 4D $\mathcal{N}=4$ SYM and $\mathcal{N}=8$ supergravity from our construction. Under the dimensional reduction, the 6D SYM superfield (\ref{eq:6DSYM}) reduces to the superfield of 4D $\mathcal{N}=4$ SYM (in the non-chiral form) as shown in (\ref{eq:non-chiral_form}). For the pair of particles that we take to be forward (i.e. the particles $(n{+}1)$ and $(n{+}2)$), we set the Grassmann variables as
\begin{equation} \label{eq:etan2}
\eta_{n+2}^1=\eta_{n+1}^1\, , \quad \eta_{n+2}^2=- \eta_{n+1}^2\, , \quad 
\tilde{\eta}_{n+2}^{\tilde 1}= \tilde{\eta}_{n+1}^{\tilde 1}, \quad  \tilde{\eta}_{n+2}^ {\tilde 2}=- \tilde{\eta}_{n+1}^{\tilde 2} \, .
\end{equation}
According to the procedure discussed in the previous section, the one-loop superamplitudes in $\mathcal{N}=4$ SYM are then given by
\begin{align} 
\mathcal{A}^{(1)}_{n}  (\alpha)
=
\int \frac{d^{4} \ell}{\ell^2} \int d\mu^{\rm 6D}_{n+2} \int d^2 \eta_{n+1} d^2 \tilde{\eta}_{n+1} \, {\cal I}_L^{(1,1)}   \, {\cal I}_R^{(n+1, \alpha, n+2)}\big{|}_{\rm F.L.} \, ,
\end{align}
with the left and right integrands given by
\bea
{\cal I}_L^{(1,1)} = \delta^{n+2} (V \cdot \Omega \cdot \eta) \delta^{n+2} (\widetilde{V} \cdot \widetilde{\Omega} \cdot
\tilde{\eta})\,  {\rm det}^{\prime} H_{n+2} \, , \quad {\cal I}_R^{(n+1, \alpha, n+2)} =\sum_{\rm cyclic} {\rm PT} (n{+}1, \alpha, n{+}2) \, ,
\eea
where the cyclic sum is to sum over all the cyclic permutation of $\alpha$. Again, it should be understood that the kinematics in the above formula are taken to be: $n$ of the helicity spinors are in 4D, and two remain in 6D and being forward, whereas the Grassmann variables are identified according to (\ref{eq:non-chiral_form}) and (\ref{eq:etan2}). 

\begin{figure}
\begin{center} 
	\begin{tikzpicture}[scale=1, line width=1 pt]
	\draw [4Dline] (0,0)--(-1.75,-1.2);
	\draw [4Dline] (2,0)--(3.75,-1.2);
	\draw [4Dline] (2,2)--(3.75,3.2);
	\draw [4Dline] (0,2)--(-1.75,3.2);
\draw[red,line width=0.05cm]
(0,0)--(0.6,-0.8);
\draw [red,line width=0.05cm]
(1.4,-0.8)--(2,0);
	\draw [6Dlinenoarrow] (2,0)--(2,2);
	\draw [6Dlinenoarrow] (2,2)--(0,2);
	\draw [6Dlinenoarrow] (0,2)--(0,0);
	\draw[black,fill=lightgray] (2,0) circle (2.5ex);
	\draw[black,fill=lightgray] (2,2) circle (2.5ex);
	\draw[black,fill=lightgray] (0,0) circle (2.5ex);
	\draw[black,fill=lightgray] (0,2) circle (2.5ex);	
		\node at (-1.9,-1.3) {$1$};
		\node at (3.9,-1.3) {$4$};
		\node at (-1.9,3.4) {$2$};
		\node at (3.9,3.4) {$3$};
		\node at (0.6,-1.1) {\color{red}{$-$}};
		\node at (1.3
		,-1.1) {\color{red}{$+$}};		
		\node at (5,1) {$\xRightarrow[\text{}]{\text{forward}}$};		
	\end{tikzpicture}
	$\quad$
	\begin{tikzpicture}[scale=1, line width=1 pt]
	\draw [4Dline] (0,0)--(-1.75,-1.2);
	\draw [4Dline] (2,0)--(3.75,-1.2);
	\draw [4Dline] (2,2)--(3.75,3.2);
	\draw [4Dline] (0,2)--(-1.75,3.2);
	\draw [6Dline] (2,0)--(0,0);
	\draw [6Dlinenoarrow] (2,0)--(2,2);
	\draw [6Dlinenoarrow] (2,2)--(0,2);
	\draw [6Dlinenoarrow] (0,2)--(0,0);
	
	\draw[black,fill=lightgray] (2,0) circle (2.5ex);
	\draw[black,fill=lightgray] (2,2) circle (2.5ex);
	\draw[black,fill=lightgray] (0,0) circle (2.5ex);
	\draw[black,fill=lightgray] (0,2) circle (2.5ex);

	\node at (1,-0.5) {$\ell$};
	\node at (-1.9,-1.3) {$1$};
	\node at (3.9,-1.3) {$4$};
	\node at (-1.9,3.4) {$2$};
	\node at (3.9,3.4) {$3$};
	\end{tikzpicture}  
\end{center}
\caption{The curvy line means the particle momentum is restricted in 4D (as parametrised in (\ref{eq:4dkin})), while the particle with the solid line still remains in 6D kinematics. The red line represents the forward-limit particle pair with 6D kinematics, which is understood as loop momentum. \label{Fig:4ptSYM}}
\end{figure}
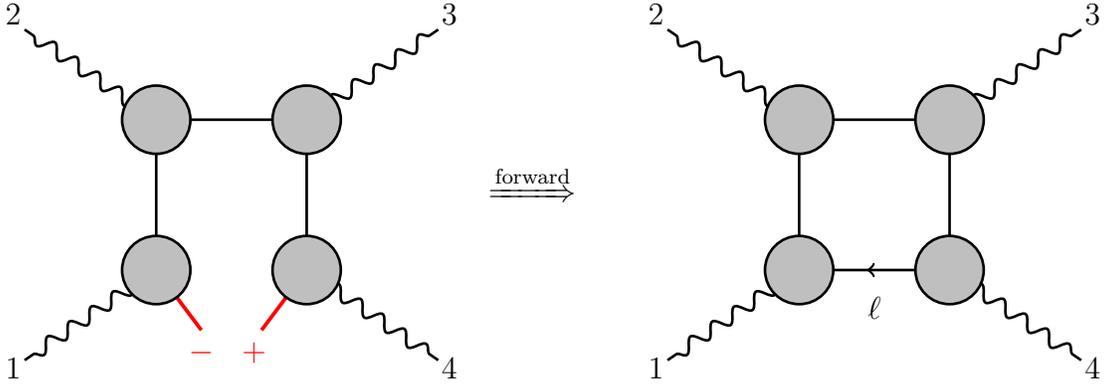

As an example, we construct the one-loop corrections to the four-point superamplitude in 4D $\mathcal{N}=4$ SYM. Following the procedure discussed previously, we begin with the six-point tree-level superamplitude of 6D $(1,1)$ SYM, and dimensionally reduce four of the external kinematics (including the fermionic ones) to 4D $\mathcal{N}=4$ SYM, whereas the remaining two of them are in 6D and are taken to be forward, as illustrated in Fig.(\ref{Fig:4ptSYM}).  We find that the one-loop integrand of the four-point amplitude in 4D $\mathcal{N}=4$ SYM is given by
\bea
\mathcal{A}_4^{(1)} =  {\delta^4(Q_4)\delta^4(\tilde{Q}_4) } \times I_{\text{box}}(k_{1},k_{2},k_{4}) \, ,
\eea
where the supercharges $Q_4, \tilde{Q}_4$ are defined in (\ref{eq:4D-charge}). The box integral $I_{\text{box}}(k_{1},k_{2},k_{4})$ is defined as
\begin{equation} \label{eq:linear-box}
I_{\text{box}}(k_{1},k_{2},k_{4})=\int d^{4} \ell\; \frac{1}{\ell^2\,\left(2\ell \cdot k_{1}\right)\left(2\ell \cdot (k_{1}+k_{2})+2k_{1}\cdot k_{2}\right)\left(-2\ell \cdot k_{4}\right)} + {\rm cyclic} \,,   
\end{equation}
where the integral is in the linear propagator representation and we also sum over four cyclic permutations. The result is in agreement with \cite{Geyer:2015bja}. Importantly, as we show in Appendix \ref{app:llinear}, the linear-propagator representation (\ref{eq:linear-box}) is equivalent to the standard box integral with quadratic propagators, namely
\bea
\tilde{I}_{\text{box}}(k_{1},k_{2},k_{4})  =  \int d^{4} \ell\; \frac{1}{\ell^2\,\left(\ell + k_{1}\right)^2 \left( \ell + k_{1}+k_{2} \right)^2 \left(\ell - k_{4}\right)^2} \, .
\eea

Similar construction applies to the one-loop corrections to the four-point superamplitude in 4D $\mathcal{N}=8$ supergravity. For the $\mathcal{N}=8$ supergravity loop amplitudes, we begin with the six-point tree-level superamplitude of $(2,2)$ supergravity. The prescription then leads to the following result 
\begin{equation} \label{eq:GrM4}
\mathcal{M}_{4}^{(1)} =  \delta^{8}(Q_4)\delta^{8}(\tilde{Q}_4)  \left[ I_{\text{box}}(k_{1},k_{2},k_{4})+\text{Perm}\right]\,,    
\end{equation}
where $Q_4, \tilde{Q}_4$ are defined in (\ref{eq:4D-charge}), but now with $I=1,2,3,4$ and $\tilde{I}=1,2,3,4$ for the $\mathcal{N}=8$ supersymmetry. We also sum over all the permutations due to the permutation symmetry of gravity amplitudes, and (\ref{eq:GrM4}) indeed reproduces the known result in \cite{Geyer:2015bja}, which is equivalent to the original result in the quadratic propagator form \cite{Green:1982sw}.  

The construction may be generalised to higher loops, especially the two-loop corrections \cite{Geyer:2016wjx, Geyer:2018xwu, Geyer:2019hnn}.  For describing two-loop corrections, 
we will need $(n{+}4)$-point tree amplitudes as input, and set two pairs of particles in forward limit ($p^{AB}_{n+1},\,p^{AB}_{n+2}$ are forward, so as $p^{AB}_{n+3},\,p^{AB}_{n+4}$), and require their supercharges to cancel among each other. To be explicit, the second pair of forward-limit particles $(n{+}3)$ and $(n{+}4)$ should obey the same relation in (\ref{eq:FD}) and (\ref{eq:FW-eta}) as the first particle pair $(n{+}1)$ and $(n{+}2)$ do. With the similar setup as one-loop amplitude, we can write down the twistor formulation of two-loop amplitude as
\begin{align} \label{eq:two-loop-twistor}
\mathcal{A}^{(2)}_{{\rm 4D}, n}&= \int \frac{d^{4} \ell_1}{\ell_1^2}\,\frac{d^{4} \ell_2}{\ell_2^2} \int \prod_{i=n+1,n+3} d^{2\mathcal{N}}\eta_{i}d^{2 \widetilde{\mathcal{N}}} \tilde{\eta}_{i}\, \mathcal{A}^{(0)}_{{\rm 6D}, n+4} \big{|}_{\rm F.L.} 
\nonumber\\
&=\int \frac{d^{4} \ell_1}{\ell_1^2}\,\frac{d^{4} \ell_2}{\ell_2^2} \int \prod_{i=n+1,n+3} d^{2\mathcal{N}}\eta_{i}d^{2 \widetilde{\mathcal{N}}} \tilde{\eta}_{i}\, \int d\mu^{\rm 6D}_{n+4}\, \mathcal{I}_L \mathcal{I}_R \big{|}_{\rm F.L.}\, ,
\end{align}
where $\ell_1$ and $\ell_2$ are the momenta of the two forward-limit particles, $(n{+}1)$ and $(n{+}3)$, which are identified as loop momenta with $\ell_1^2\neq 0$ and $\ell_2^2\neq 0$. We denote the two-loop $n$-point amplitude in 4D as $\mathcal{A}^{(2)}_{{\rm 4D}, n}$. The internal states of all forward-limit particles are summed over by performing Grassmann integral. 

Using the formula (\ref{eq:two-loop-twistor}), we reproduce the well-known four-point two-loop SYM and supergravity amplitudes in \cite{Bern:1997nh, Bern:1998ug, Geyer:2016wjx, Geyer:2019hnn} \footnote{In the work of \cite{Geyer:2019hnn}, the two-loop scattering equations have two different choices, $\alpha=\pm 1$ (here $\alpha$ refers to the parameter in their paper, not the colour ordering), which corresponds to our choice of setting the momenta to be $2\, p_{n+1} \cdot p_{n+3}=\alpha\,(\ell_1+\alpha \, \ell_2)^2$. Here we have chosen $\alpha=1$, and we could have made the choice with $\alpha=-1$ to obtain the same result.}. 
The two-loop SYM can be obtained by using the 6D $\mathcal{N}=(1,1)$ SYM expressed in twistor formula
\begin{align} \label{eq:two-loopYM}
\mathcal{A}^{(2)}_{n}  (\alpha)
=
\int d\mu^{\rm 6D}_{n+4} \int \prod_{i=n+1,n+3} d^2 \eta_{i} d^2 \tilde{\eta}_{i} \, {\cal I}_L^{(1,1)}   \, {\cal I}_R^{(\text{2-loop}, \alpha)} \big{|}_{\rm F.L.} \, ,
\end{align}
where the left integrand with $(1,1)$ supersymmetry is 
\be
{\cal I}_L^{(1,1)} = \delta^{n+4} (V \cdot \Omega \cdot \eta) \delta^{n+4} (\widetilde{V} \cdot \widetilde{\Omega} \cdot
\tilde{\eta})\,  {\rm det}^{\prime} H_{n+4} \,,
\ee 
and the right integrand is a two-loop Parke-Taylor factor, which in general contains the planar and non-planar parts of the SYM amplitude. For example, the planar part is given by the following integrand
\begin{align} \label{eq:two-loopPT}
{\cal I}_R^{(\text{2-loop}, \alpha),{\rm P}}=\; &c^{\rm P}_{\a}(n{+}1,n{+}2,n{+}3,n{+}4) +c^{\rm P}_{\a}(n{+}2,n{+}1,n{+}4,n{+}3)\nonumber \\
+& c^{\rm P}_{\a}(n{+}3,n{+}4,n{+}1,n{+}2) +c^{\rm P}_{\a}(n{+}4,n{+}3,n{+}2,n{+}1)\,,
\end{align}
where the superscript ${\rm P}$ denotes planar, and the factor $c^{\rm P}_{\a}(a,b,c,d)$ is defined as
\begin{align}
c^{\rm P}_{\a}(a,b,c,d)=\sum_{{\rm cyclic}\; \a}\Big(&{\rm PT}(a,b,d,c,\a_1,\a_2,\a_3,\a_4)+{\rm PT}(a,b,\a_1,d,c,\a_2,\a_3,\a_4)\nonumber\\
+&{\rm PT}(a,b,\a_1,\a_2,d,c,\a_3,\a_4)+{\rm PT}(a,c,d,b,\a_1,\a_2,\a_3,\a_4)\Big) \,.
\end{align}
We have checked that (\ref{eq:two-loopYM}) gives correct two-loop four-point amplitude; for instance,
the planar part computed by using the integrand in (\ref{eq:two-loopPT}) is shown to be equal to
\begin{align}
\label{eq:two-loopP}
\mathcal{A}^{(2),{\rm P}}(\a)=\delta^4(Q_4)\delta^4(\tilde{Q}_4) \Big(\,k_{\a_1}\cdot k_{\a_2}\; I^{\text{planar}}_{\a_1\a_2,\,\a_3\a_4} +k_{\a_4}\cdot k_{\a_1}\; I^{\text{planar}}_{\a_4\a_1,\,\a_2\a_3}\Big)\, , 
\end{align}
where the planar two-loop boxes, $I^{\text{planar}}_{\a_1\a_2,\,\a_3\a_4}$ are defined in the Appendix.A of \cite{Geyer:2016wjx}, see equation (A5). We have also checked the agreement between our formula and the known result for the non-planar sector.  Similarly, the four-point two-loop superamplitude in 4D $\mathcal{N}=8$ supergravity can be obtained by considering the eight-point tree-level amplitude of the 6D $(2,2)$ supergravity, which is expressed in the twistor formulation as given in (\ref{eq:22SUGRA}). We have verified that the result of this construction is in the agreement with the known result \cite{Geyer:2016wjx, Bern:1998ug}. 
\section{Supersymmetric D3-brane amplitudes at one loop} 
\label{sec:D3loop}

In this section, we consider one-loop corrections to the superamplitudes in the D3-brane theory using forward limits of higher-dimensional tree-level amplitudes, following the general prescription of the previous sections. The results will be further confirmed using generalised unitarity methods. The higher-dimensional amplitudes that are relevant for constructing the loop corrections to D3-brane amplitudes are the tree-level amplitudes in the M5-brane theory\footnote{The same results can be obtained if we use the D5-brane tree amplitudes. The D5-brane theory has different supersymmetry from the M5-brane theory; however, such difference no longer exists after dimensional reduction to 4D. We have checked explicitly for a few examples that the lower-dimensional results are indeed independent of the choices. In practice, the twistor formula for M5-brane tree amplitudes is simpler, which involves only the ``left-handed" variables since it is a chiral theory, as can be seen from (\ref{D5-R}) and (\ref{M5-R}).}.  Because the construction makes the supersymmetry manifest, the amplitudes with all-plus and single-minus helicity configurations manifestly vanish. So the first non-trivial helicity configurations are those with two minus photons, namely the MHV amplitudes.

It is known that the non-trivial tree-level amplitudes in the D3-brane theory are helicity conserving~\cite{Rosly:2002jt}. So the MHV amplitudes vanish at tree level (except for the four-point case), and they do not have non-trivial four-dimensional cuts at one-loop order. Therefore the one-loop MHV amplitudes (with more than four points) can only be rational terms. To extract the rational terms, it is necessary to consider the loop momenta in general $d$ dimensions. We will treat the extra-dimensional loop momenta as masses, therefore, it is equivalent to consider 4D loop momenta with massive states running in the loop.

In our construction, this is set up by separating the massless 6D momenta (which will be taken to be forward) into $\ell_{\rm 4D}$ and two extra dimensions, such that when the loop momentum $\ell_{\rm 4D}$ is put on-shell, we have $\ell_{\rm 4D}^2 = - \mu^2$.  
With such convention, we write a 6D momentum as its 4D component ($\ell_{\rm 4D}:=\ell$)
with extra dimensions, that will be called as $p_4$ and $p_5$, which correspond to the components of the fifth and the sixth dimension, respectively. 
The 6D momentum is explicitly written as
\be
\ell_{\rm 6D}=(\ell,p_4,p_5)\, , 
\ee
then the massless condition is
\be  \label{eq:mu}
0=\ell_{\rm 6D}^2=\ell^2+\mu\tilde{\mu}\,,  \quad {\rm with} \quad  \mu=p_4+ip_5,\quad \tilde{\mu}=p_4-ip_5\,.
\ee
Since we have identified the momentum of a forward-limit particle as the loop momentum as shown in (\ref{eq:FD}), $\mu$ and $\tilde{\mu}$ can also be expressed in terms of $\lambda_{n+1,a}^{A}$:
\be
\mu=\langle \lambda_{n+1}^1 \, \lambda_{n+1}^{2} \rangle,\qquad \tilde{\mu}= \langle \lambda_{n+1}^3\, \lambda_{n+1}^{4} \rangle \,.
\ee
We now have a massive particle in the loop with a loop momentum, $\tilde{\ell}=\ell+\mu$ with $\tilde{\ell}^2=\ell^2+\mu^2$. Also, the linear propagators are unchanged since $\mu$ is in the extra dimension, so we have
\be
\tilde{\ell} \cdot k_i= (\ell+\mu) \cdot k_i=\ell \cdot k_i\, ,
\ee
for a 4D external momentum $k_i$. 

With this setup, the one-loop D3-brane amplitudes can be obtained from the tree-level M5-brane amplitude by a similar construction we outlined in previous sections, 
\be \label{eq:D3loop}
\mathcal{A}^{(1)}_{\text{D3},\,n} = \int \frac{d^D \tilde{\ell}}{(2\pi)^D}\, \frac{1}{\tilde{\ell}^2}\, \int d^4\eta_{n+1} \, \mathcal{A}^{(0)}_{\text{M5},\,n+2} \big{|}_{\rm F.L.} \,,
\ee
where $d^4\eta_{n+1}=d\eta^1_{n+1,1}d\eta^2_{n+1,1}d\eta^1_{n+1,2}d\eta^2_{n+1,2}$, and Grassmann integration is to sum over the superstates of the forward-limit particles. Again in the above formula, it should be understood that it is the tree-level forward-limit amplitude on the right-hand side that gives the one-loop integrand.  This tree-level M5-brane amplitude is expressed in the twistor formulation as
\be
\mathcal{A}^{(0)}_{\text{M5},\,n+2}= \int  d\mu^{\rm 6D}_{n+2}\; \mathcal{I}^{\text{M5}}_L \, \mathcal{I}^{\text{M5}}_R \, ,
\ee 
where the left and right integrands  $ \mathcal{I}^{\text{M5}}_L$ and $\mathcal{I}^{\text{M5}}_R $ are given in (\ref{M5-L}), and (\ref{M5-R}), respectively.

\subsection{One-loop corrections to D3-brane superamplitudes}

\begin{figure}
\begin{center}
\begin{tikzpicture}[scale=1, line width=1 pt]
		\draw [4Dline] (-1.5,1.2)--(0,0);
		\draw [4Dline] (-1.5,-1.2)--(0,0);
		\draw [6Dlinenoarrow] (0,0) arc (150:30:1.15);
		
		\draw [red,line width=0.05cm] (2,0) arc (0:-50:1.25);
		\draw [red,line width=0.05cm] (0,0) arc (-180:-130:1.25);
		
     	\draw [4Dline] (2,0)--(3.5,1.2);
		\draw [4Dline] (2,0)--(3.5,-1.2);
		\draw[black,fill=lightgray] (0,0) circle (2.35ex);
		\draw[black,fill=lightgray] (2,0) circle (2.35ex);
		\node at (0,0){$Q_{K_1}$};
		\node at (2,0) {$Q_{K_2}$};
		\node at (-2,-1.2){$1$};
		\node at (-2,1.2) {$2$};
		\node at (4,1.2) {$3$};
		\node at (4,-1.2) {$4$};
		
		\node at (1.30,-1) {{\color{red}{$+$}}};
		\node at (0.75,-1) {{\color{red}{$-$}}};
		\node at (4.8,0) {$\xRightarrow[\text{}]{\text{forward}}$};
		\end{tikzpicture}
		\begin{tikzpicture}[scale=1, line width=1 pt]
		\draw [4Dline] (-1.5,1.2)--(0,0);
		\draw [4Dline] (-1.5,-1.2)--(0,0);
		\draw [6Dlinenoarrow] (0,0) arc (150:30:1.15);
		\draw [6Dline] (2,0) arc (-20:-155:1.05);
		
     	\draw [4Dline] (2,0)--(3.5,1.2);
		\draw [4Dline] (2,0)--(3.5,-1.2);
		\draw[black,fill=lightgray] (0,0) circle (2.35ex);
		\draw[black,fill=lightgray] (2,0) circle (2.35ex);
		\node at (0,0){$Q_{K_1}$};
		\node at (2,0) {$Q_{K_2}$};
		\node at (-2,-1.2){$1$};
		\node at (-2,1.2) {$2$};
		\node at (4,1.2) {$3$};
		\node at (4,-1.2) {$4$};
		\node at (1,-1.2)
		{$\ell$};
		\end{tikzpicture}
		\label{fig:bubble}
	\caption{The bubble diagram comes from gluing the leg-$(n{+}1)$ and leg-$(n{+}2)$ (denoted by $-$ and $+$, respectively) of a six-point tree amplitude. The original $s_{1\,2\,n+1}$ channel becomes the linear propagator of the first term in (\ref{eq:bubble}).}
\end{center}
\end{figure}
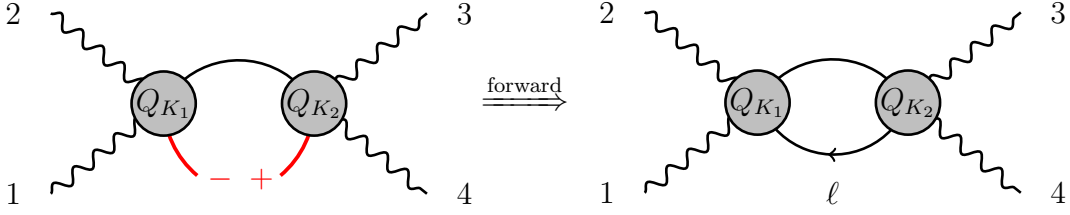

Let us begin with the one-loop correction to a four-point amplitude, which is very similar to the case of four-point one-loop amplitudes of $\mathcal{N}=4$ SYM in the previous section. We will show that the one-loop amplitude receives correction from the bubble diagram in the Fig.(\ref{fig:bubble}). Using the general formula (\ref{eq:D3loop}) and solving the scattering equations, we find that the one-loop correction to the four-point superamplitude in the D3-brane theory takes the following form
\begin{equation} \label{eq:D3loop_4pt}
\mathcal{A}^{(1)}_{ \text{D3}\,,4}=\mathcal{A}^{(0)}_{\text{D3}\,,4}
\left(s_{12}^2\, I_{\text{bubble}}(k_{1},k_{2})+\text{Perm}\right)\,,    
\end{equation}
where $\mathcal{A}^{(0)}_{\text{D3}\,,4}$ is the tree-level amplitude of the D3-brane theory, and it is given by
\be
\mathcal{A}^{(0)}_{\text{D3}\,,4}=
\delta^{4}(\sum_{i=1}^4 \lambda^{\alpha}_i\eta_{i,-}^I)
\delta^{4}(\sum_{i=1}^4\tilde{\lambda}^{\dot{\alpha}}_i\eta_{i,+}^{\tilde{I}}) \, .
\ee
Note that it is the supersymmetrisation of the higher-derivative term $F^4$.  As we have seen in the previous section, in this construction the loop integrands are typically in the linear propagator representation. For the four-point case we consider here, the bubble integral is defined by
\begin{align} \label{eq:bubble}
I_{\text{bubble}}(k_{1},k_{2})=\int \frac{d^{D} \tilde{\ell} }{(2\pi)^{D}}\, &\Big[ \;\frac{1}{\tilde{\ell}^2 \,(2\tilde{\ell} \cdot (k_{1}+k_{2})+2k_{1}\cdot k_{2})}\nonumber\\
&+\frac{1}{\tilde{\ell}^2 \,(-2 \tilde{\ell} \cdot (k_{1}+k_{2})+2k_{1}\cdot k_{2})}\Big] \, .
\end{align}
Recall $\tilde{\ell} = \ell+ \mu$, so that $\tilde{\ell}^2 = \ell^2 + \mu^2$ and $\tilde{\ell} \cdot (k_{1}+k_{2}) ={\ell} \cdot (k_{1}+k_{2})$. As shown in Appendix \ref{app:llinear},  (\ref{eq:bubble}) is equivalent to the standard bubble integral with quadratic propagators
\begin{align} \label{eq:bubble-2}
\tilde{I}_{\text{bubble}}(k_{1},k_{2})=\int \frac{d^{D} \tilde{\ell} }{(2\pi)^{D}}\, &\Big[ \;\frac{1}{\tilde{\ell}^2 \, (\tilde{\ell} + k_{1}+k_{2} )^2} \Big] \, . 
\end{align}
The result (\ref{eq:D3loop_4pt}) is in agreement with the result in the reference \cite{Shmakova:1999ai} that was originally obtained using unitarity cuts~\cite{Bern:1994zx, Bern:1994cg}, which is in the quadratic propagator form (\ref{eq:bubble-2}), and only massless loop propagators were considered.  The bubble integral (\ref{eq:bubble}) or (\ref{eq:bubble-2}) is UV divergent in 4D, therefore the result (\ref{eq:D3loop_4pt}) leads to a UV counter term for the D3-brane effective Lagrangian, which is of the form $d^4F^4$ (and its supersymmetric completion), or equivalently in momentum space it is given as
\bea
(s_{12}^2+s_{23}^2+ s_{13}^2)\, \delta^{4}(\sum_{i=1}^4 \lambda^{\alpha}_i\eta_{i,-}^I)
\delta^{4}(\sum_{i=1}^4\tilde{\lambda}^{\dot{\alpha}}_i\eta_{i,+}^{\tilde{I}}) \, .
\eea 

The above four-point superamplitude is expressed in a non-chiral form, where the superfield is given by (\ref{eq:non-chiral_form}). For describing the MHV superamplitudes at higher points which we will study shortly, it is more convenient to use the chiral version. The chiral version is obtained by a Grassmann Fourier transform. For instance, the chiral superfield is obtained from non-chiral superfield given in (\ref{eq:non-chiral_form}) through, 
\be 
V_{\mathcal{N}=4}(\eta_-,  \xi)  =  \int d^2 \eta_+ e^{ \eta^{\tilde I}_+ \xi_{\tilde I}}  \Phi_{\mathcal{N}=4}(\eta_+, {\eta}_-) \, .
\ee 
After combining $\eta^I_-$ and $\xi_{\tilde I}$ and denoting them as $\eta^A$ with $A=1,2,3,4$, which transform under $SU(4)$ R-symmetry, we obtain the superfield in a chiral form,
\begin{equation} 
V_{\mathcal{N}=4}(\eta) = A^{+} + \eta^A \psi^+_A +  \eta^A \eta^B \phi_{AB}
+ \frac{1}{3!}\e_{ABCD}\eta^A \eta^B \eta^C \psi^{D -}
+ \eta^1\eta^2\eta^3\eta^4 A^{-}\, ,
\end{equation}
and the supercharges take the form
\begin{equation} \label{4dsupercharges}
q^{\a A}_i = \l^\a_i \eta^A_i \quad {\rm and} \quad q^{\dot\a}_{i, A}
= \tilde\l^{\dot\a}_i \frac{\pa}{\pa \eta^A_i}  \, . 
\end{equation}
In the chiral representation, the four-point tree-level D3-brane superamplitude is given by \cite{Chen:2015hpa}
\begin{equation} \label{A4D3b}
\mathcal{A}^{(0)}_{\text{D3}\,,4} = \left( \frac{[12]}{\langle 34 \rangle }\right)^2  
\,
\d^8 \left(\sum_{i=1}^4 q_i^{\a I} \right),
\end{equation}
where $ \left( \frac{[12]}{\langle 34 \rangle }\right)^2$ is the Jacobian factor from the Grassmann Fourier transform, and importantly it is permutation invariant and $\langle 34 \rangle^2$ in the denominator is not a pole. When expressed in the chiral form, the four-point D3-brane superamplitude at one loop takes the following form
\begin{equation} \label{eq:D3loop_4pt_chiral}
\mathcal{A}^{(1)}_{\text{D3}\,,4}=\d^8 \left( Q_4 \right) \left( [12]^2[34]^2 \times I_{\text{bubble}}(k_{1},k_{2})+\text{Perm}\right)\, ,
\end{equation}
where we define $Q^{\a A}_n = \sum_{i=1}^n q_i^{\a A}$. We will continue to utilise the chiral representation for the discussion on higher-point D3-brane superamplitudes.

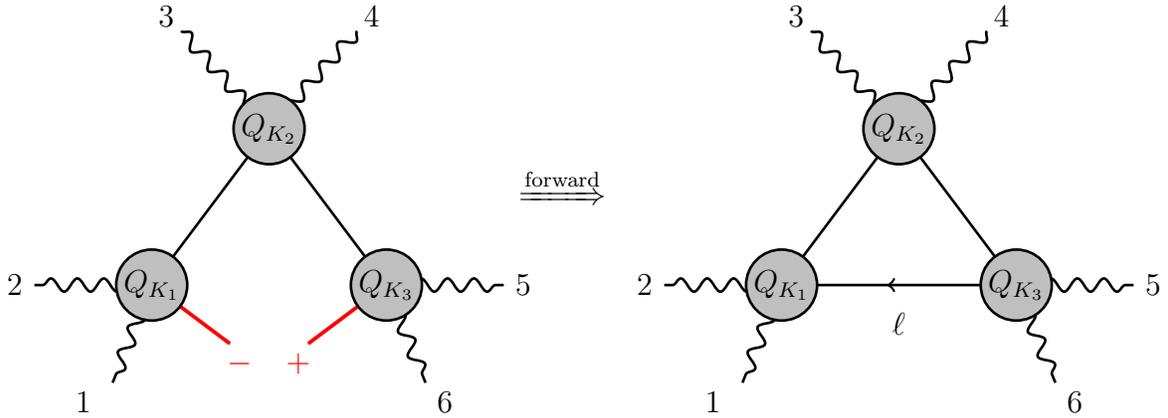
\begin{figure}
\begin{center}
{\begin{tikzpicture}[scale=1.3, line width=1 pt]
		\draw [4Dline] (-0.9,1.2)--(0,0.2);
		\draw [4Dline] (0.9,1.2)--(0,0.2);
		\draw [6Dlinenoarrow] (-1.2,-1.4)--(0,0.2);
		\draw [6Dlinenoarrow] (0,0.2)--(1.2,-1.4);
		\draw [4Dline] (-1.2,-1.4)--(-2.4,-1.4);
		\draw [4Dline] (-1.2,-1.4)--(-1.6,-2.4);
		\draw [4Dline] (1.2,-1.4)--(2.4,-1.4);
		\draw [4Dline] (1.2,-1.4)--(1.6,-2.4);
		\draw [red,line width=0.05cm]
        (-1.2,-1.4)--(-0.4,-2);
        \draw [red,line width=0.05cm]
        (1.2,-1.4)--(0.4,-2);
        
		\node at (-0.3,-2.2) {{\color{red}{$-$}}};
		\node at (0.3,-2.2) {{\color{red}{$+$}}};
		\node at (-1.05,1.35) {$3$};
		\node at (1.05,1.35) {$4$};
		\node at (-1.9,-2.6) {$1$};
		\node at (1.8,-2.6) {$6$};
		\node at (-2.6,-1.4) {$2$};
		\node at (2.6,-1.4) {$5$};
		\draw[black,fill=lightgray] (0,0.2) circle (2ex);
		\node at (0,0.2) {$Q_{K_2}$};
		\draw[black,fill=lightgray] (-1.2,-1.4) circle (2ex);
		\node at (-1.2,-1.4) {$Q_{K_1}$};
		\draw[black,fill=lightgray] (1.2,-1.4) circle (2ex);
		\node at (1.2,-1.4) {$Q_{K_3}$};
		\node at (3,-0.5)
		{$\xRightarrow[\text{}]{\text{forward}}$};
		\end{tikzpicture}}
{\begin{tikzpicture}[scale=1.3, line width=1 pt]
		\draw [4Dline] (-0.9,1.2)--(0,0.2);
		\draw [4Dline] (0.9,1.2)--(0,0.2);
		\draw [6Dlinenoarrow] (-1.2,-1.4)--(0,0.2);
		\draw [6Dlinenoarrow] (0,0.2)--(1.2,-1.4);
		\draw [6Dline] (1.2,-1.4)--(-1.2,-1.4);
		\draw [4Dline] (-1.2,-1.4)--(-2.4,-1.4);
		\draw [4Dline] (-1.2,-1.4)--(-1.6,-2.4);
		\draw [4Dline] (1.2,-1.4)--(2.4,-1.4);
		\draw [4Dline] (1.2,-1.4)--(1.6,-2.4);
		
		\node at (0,-1.8) {$\ell$};
		\node at (-1.05,1.35) {$3$};
		\node at (1.05,1.35) {$4$};
		\node at (-1.9,-2.6) {$1$};
		\node at (1.8,-2.6) {$6$};
		\node at (-2.6,-1.4) {$2$};
		\node at (2.6,-1.4) {$5$};
		\draw[black,fill=lightgray] (0,0.2) circle (2ex);
		\node at (0,0.2) {$Q_{K_2}$};
		\draw[black,fill=lightgray] (-1.2,-1.4) circle (2ex);
		\node at (-1.2,-1.4) {$Q_{K_1}$};
		\draw[black,fill=lightgray] (1.2,-1.4) circle (2ex);
		\node at (1.2,-1.4) {$Q_{K_3}$};
		\end{tikzpicture}}
\caption{Triangle diagram arises from gluing an eight-point tree diagram, where the $s_{1\,2\,n+1}$ and $s_{5\,6\,n+2}$ channels give rise to the linear propagators in the first term of (\ref{eq:6pt_rational}). The other two terms correspond to identifying the other two internal lines in the diagram with loop momentum. }
\label{fig:6pt_triangle}
\end{center}
\end{figure}

The one-loop corrections to higher-point superamplitudes can be obtained similarly from the general formula (\ref{eq:D3loop}). To illustrate the idea, we will consider the six-point MHV amplitude. Using the formula (\ref{eq:D3loop}) with $n=6$, we find the MHV amplitude is proportional to an overall supercharge, $\delta^8(Q_6)$, as required by $\mathcal{N}=4$ supersymmetry. Explicitly, the one-loop superamplitude is constructed by taking forward the eight-point tree-level superamplitude as shown in the Fig.(\ref{fig:6pt_triangle}). Note that the contact term of eight-point amplitude does not contribute. We find that the integrand for the one-loop six-point MHV amplitude takes a very similar form as the four-point result given in (\ref{eq:D3loop_4pt_chiral}). It is given as
\begin{align} \label{eq:6pt_rational}
\mathcal{A}^{(1),\text{MHV}}_{\text{D3}\,,6} = & \,\, 
\delta^8(Q_6) \left(\mu^2\, [12]^2[34]^2[56]^2 \times I_{\rm triangle} (k_1,k_2; k_3, k_4; k_5, k_6)  +\text{Perm}\right)  \, .
\end{align}
The integral $I_{\rm triangle}$ is the scalar triangle integral in the linear propagator representation, which takes the following form
\begin{align} \label{eq:6pt_rational2}
 I_{\rm triangle}  (k_1,k_2; k_3, k_4; k_5, k_6) &=  -  \int \frac{d^D\tilde{\ell}}{(2\pi)^D}\Big(\; \frac{1}{\tilde{\ell}^2\, [2\tilde{\ell} \cdot (k_{1}+k_{2})+2k_1\cdot k_2]\,[-2\tilde{\ell} \cdot (k_5+k_6)+2k_5 \cdot k_6]}\nonumber\\
&~~~ +\frac{1}{ \tilde{\ell}^2\,  [2\tilde{\ell} \cdot (k_{3}+k_{4})+2k_3\cdot k_4]\,[-2\tilde{\ell} \cdot (k_1+k_2)+2k_1 \cdot k_2]} \nonumber \\
&~~~ +\frac{1}{\tilde{\ell}^2  \, [2\tilde{\ell} \cdot (k_{5}+k_{6})+2k_5\cdot k_6]\,[-2\tilde{\ell} \cdot (k_3+k_4)+2k_3 \cdot k_4]}\Big)  \, . 
\end{align}
This result is verified by (\ref{eq:D3loop}) for $n=6$ by solving numerically the scattering equations in the formula with the forward kinematics. There are three terms with linear propagators in the above equation, each of them can be understood as assigning the forward pair of legs in different places; for example, the first term in (\ref{eq:6pt_rational}) is shown in the Fig.(\ref{fig:6pt_triangle}).  In principle, the one-loop corrections to higher-point amplitudes can be obtained in a similar fashion; however, solving scattering equations with higher-point  kinematics becomes more and more difficult. We hope to develop better numerical and analytical methods to handle this issue, which we leave as a further research direction.

\subsection{Generalised unitarity methods}
\label{sec:GUM}

In this section, we will construct the one-loop amplitudes through the $d$-dimensional generalised unitary methods~\cite{Anastasiou:2006jv}. The  $d$-dimensional  cuts are necessary to extract the rational terms of the loop amplitudes since the rational terms have vanishing four-dimensional cuts. As in the previous section, again the extra dimensional loop momenta will be viewed as the masses of the internal propagating particles in 4D. Therefore, effectively we will perform four-dimensional cuts, but with massive loop momenta.   We will find the results agree with those computed in the previous section using the twistor formulations. Of course, they are in different representations: one in the linear propagator representation, the other in the standard quadratic propagator representation.

\begin{figure}
\begin{center}
		\begin{tikzpicture}[scale=1, line width=1 pt]
		\draw [4Dline] (-1.5,1.2)--(0,0);
		\draw [4Dline] (-1.5,-1.2)--(0,0);
		\draw [6Dlinenoarrow] (0,0) arc (150:30:1.15);
		\draw [6Dlinenoarrow] (2,0) arc (-20:-155:1.05);
		\draw [dashed,gray] (1,0.8)--(1,0.3);
		\draw [dashed,gray] (1,-0.9)--(1,-0.3);
     	\draw [4Dline] (2,0)--(3.5,1.2);
		\draw [4Dline] (2,0)--(3.5,-1.2);
		\draw[black,fill=lightgray] (0,0) circle (2.5ex);
		\draw[black,fill=lightgray] (2,0) circle (2.5ex);
		\node at (0,0){$Q_{K_1}$};
		\node at (2,0) {$Q_{K_2}$};
		\node at (-2,-1.2){$1$};
		\node at (-2,1.2) {$2$};
		\node at (4,1.2) {$3$};
		\node at (4,-1.2) {$4$};
		\node at (1,-1.3)
		{$K_1$};
		\node at (1,1.2)
		{$K_2$};
		\end{tikzpicture}
		\label{fig:4pt_bubble_gluing}
	\caption{The four-point bubble diagram is formed by gluing two tree-level four-point superamplitudes. We identify $K_1$ as loop momentum ($K_1=\ell$), then the on-shell conditions ($K_i^2=0$) can be read as $\ell^2=0$, and $(\ell+k_1+k_2)^2=0$.}
\end{center}
\end{figure}
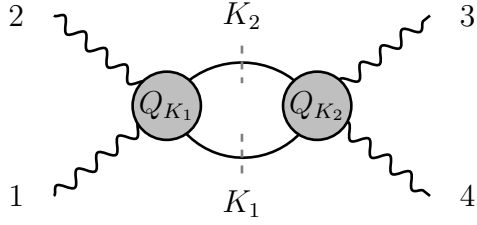

Let us first begin with the four-point case. The one-loop amplitude only receives contribution from the bubble diagram, which can be formed by gluing two four-point superamplitudes as shown in the Fig.(\ref{fig:4pt_bubble_gluing}).
Explicitly, it is given by 
\be  \label{eq:glue_bubble_4pt}
\int \frac{d^D\tilde{\ell}}{(2\pi)^D} \int \big(
\prod_{i=1}^{2}d^4\eta_{K_i}\big) \big(\prod_{i=1}^2 \frac{1}{K_i^2}\big)\, \delta^{8}(Q_{K_1})\delta^{8}(Q_{K_2})\,,
\ee
where the explicit form of the four-point superamplitudes are given by 
\be
Q_{K_1}=\sum_{i=1,2,K_{1},-K_{2}}\lambda^A_{i,a}\eta_{i}^{I,a}\,, \quad \text{and} \quad  Q_{K_2}=\sum_{i=3,4,K_{2},-K_{1}} \lambda^A_{i,a}\eta_{i}^{I,a}\, .
\ee
Note for the external states, they are massless, therefore 
\begin{equation} 
\lambda_{i,a}^{A}= \left( \begin{array}{cccc}
0  &\lambda^{\alpha}_i \\
\tilde{\lambda}^{\dot{\alpha}}_i  & 0
\end{array}  \right), \quad \text{for}\;\; i=1,...,4 \, . 
\end{equation}
From (\ref{eq:glue_bubble_4pt}), we deduce that the one-loop correction to the MHV amplitude is given by
\be \label{eq:glue_triangle_MHV}
\delta^{8}(Q_4)\, [12]^2[34]^2  \int \frac{d^D\tilde{\ell}}{(2\pi)^D}\,
\left( \frac{1}{{\tilde{\ell}}^2(\tilde{\ell}+k_1+k_2)^2} \right)  + {\rm Perm}\, .
\ee
In the above formula we have summed over the permutations, and after linearising the propagators as we show in Appendix \ref{app:llinear}, the result agrees with the one obtained from the twistor formula given in (\ref{eq:D3loop_4pt_chiral}).

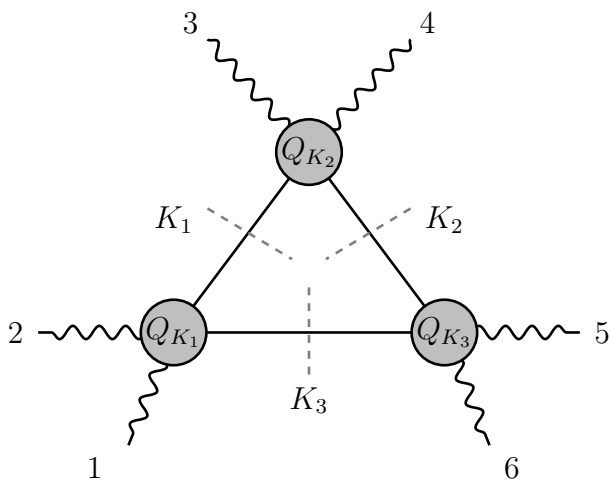
\begin{figure}
\begin{center}
	{\begin{tikzpicture}[scale=1.5, line width=1 pt]
		\draw [4Dline] (-0.9,1.2)--(0,0.2);
		\draw [4Dline] (0.9,1.2)--(0,0.2);
		\draw [6Dlinenoarrow] (-1.2,-1.4)--(0,0.2);
		\draw [6Dlinenoarrow] (1.2,-1.4)--(-1.2,-1.4);
		\draw [6Dlinenoarrow] (0,0.2)--(1.2,-1.4);
		\draw [4Dline] (-1.2,-1.4)--(-2.4,-1.4);
		\draw [4Dline] (-1.2,-1.4)--(-1.6,-2.4);
		\draw [4Dline] (1.2,-1.4)--(2.4,-1.4);
		\draw [4Dline] (1.2,-1.4)--(1.6,-2.4);
		\draw [dashed,gray] (0,-1)--(0,-1.8);
		\draw [dashed,gray] (-0.9,-0.3)--(-0.15,-0.75);
		\draw [dashed,gray] (0.9,-0.3)--(0.15,-0.75);
		\node at (-1.05,1.35) {$3$};
		\node at (1.05,1.35) {$4$};
		\node at (-1.9,-2.6) {$1$};
		\node at (1.8,-2.6) {$6$};
		\node at (-2.6,-1.4) {$2$};
		\node at (2.6,-1.4) {$5$};
			\node at (0,-2) {$K_3$};
			\node at (-1.2,-0.4) {$K_1$};
			\node at (1.2,-0.4) {$K_2$};
		\draw[black,fill=lightgray] (0,0.2) circle (1.6ex);
		\node at (0,0.2) {$Q_{K_2}$};
		\draw[black,fill=lightgray] (-1.2,-1.4) circle (1.6ex);
		\node at (-1.2,-1.4) {$Q_{K_1}$};
		\draw[black,fill=lightgray] (1.2,-1.4) circle (1.6ex);
		\node at (1.2,-1.4) {$Q_{K_3}$};
		\end{tikzpicture}}
\caption{Legs-(1,2), (3,4), and (5,6) are glued in three different corners with on-shell propagators, $K_1$, $K_2$ and $K_3$. When $K_3$ is identified with the loop momentum, the diagram gives the contribution as (\ref{eq:glue_triangle_MHV}).  } \label{fig:triangle}
\end{center}
\end{figure}

We now consider the six-point amplitude. Due to the fact that there is no six-point contact term, for the MHV amplitude, only the triangle diagram is non-trivial, for which we glue three four-point superamplitudes as shown in the Fig.(\ref{fig:triangle}). We pair up external leg-$(1,2)$, $(3,4)$, and $(5,6)$ in three different corners, and the internal massive lines are denoted as $K_{1}$, $K_2$, and $K_3$. The suprsymmetric gluing result of four-point superamplitudes gives 
\be \label{eq:glue_triangle}
\int \frac{d^D\tilde{\ell}}{(2\pi)^D} \int  \big(
\prod_{i=1}^{3}d^4\eta_{K_i}\big) \big(\prod_{i=1}^3 \frac{1}{K_i^2}\big)\, \delta^{8}(Q_{K_1})\delta^{8}(Q_{K_2})\delta^{8}(Q_{K_3})\, .
\ee
Explicit evaluation of the above formula leads to the following result for the one-loop correction to the six-point MHV amplitude
\be \label{eq:glue_triangle_MHV}
- \delta^{8}(Q_6)\, [12]^2[34]^2[56]^2  \int \frac{d^D\tilde{\ell}}{(2\pi)^D}\,
\left( \frac{\mu^2}{{\tilde{\ell}}^2(\tilde{\ell}+k_1+k_2)^2\,(\tilde{\ell}-k_5-k_6)^2} \right) +{\rm Perm} \, .
\ee
Here we have chosen the loop momentum to be $K_3$ ($K_3=\tell$), and we have also summed over permutations to obtain the complete answer.  Again, as explained in Appendix \ref{app:llinear}, (\ref{eq:glue_triangle_MHV}) is in agreement with (\ref{eq:6pt_rational}), which we obtained from the twistor formulations.  

As for a general $n$-point MHV amplitude, it is easy to see that it contains a $(n/2)$-gon integrand which takes the same form as the four- and six-point amplitudes we have computed, namely, 
\begin{align} \label{eq:npts}
 \delta^{8}(Q_n) \int \frac{d^D\tilde{\ell}}{(2\pi)^D} \Big[\frac{ (-1)^{n/2} \mu^{2(\frac{n}{2}-2)}\;[12]^2[34]^2 \cdots [n{-}1\,n]^2}{ \tilde{\ell}^2\,  (\tilde{\ell} + \sum_{i=1}^{2}k_i)^2(\tilde{\ell} + \sum_{i=1}^{4}k_i)^2
\cdots (\tilde{\ell} +\sum_{i=1}^{n-2}k_i)^2 } +\text{Perm}\Big] \, . 
\end{align}
However, as we know that for general helicity configurations, the tree-level amplitudes (with two massive states) at higher points require contact terms, (e.g. (\ref{eq:A8-tree}) for the eight-point case), it implies that besides the $({n/2})$-gon topology, the one-loop amplitude in general also receives contributions from the integrands with lower-gon topologies. For instance, for the eight-point MHV amplitude, a bubble diagram will also contribute due to the contact term shown in (\ref{eq:A8-tree}) (as well as its supersymmetric completion). We leave the computation of the one-loop corrections to the $n$-point MHV amplitude as a future research direction.

\subsection{Rational terms of MHV D3-brane amplitudes at one loop}
\label{sec:nptRationalTerm}

In the previous sections, we obtained the one-loop integrand for the MHV amplitude of the D3-brane theory, either using the forward limit of the twistor formulations or the generalised unitary methods. The one-loop integral can be performed explicitly using the dimension shifting formula \cite{Bern:1995db}, see for instance the Appendix C of \cite{Elvang:2019twd}. Explicitly, for a $m$-gon scalar integral, we have
\begin{align}  \label{eq:integral}
 \int \frac{d^{D}\tell}{(2\pi)^{D}} \frac{\mu^{2(m-2)}}{\prod_{i=1}^{m}(\tell + \sum_{j=1}^{i}K_j)^2}  =  -\frac{i(-1)^m}{(4\pi)^{2}}\frac{\Gamma(m-2)}{\Gamma(m)} + \mathcal{O}(\epsilon) \, ,
\end{align} 
where we have taken $D=4-2\epsilon$ and considered the small $\epsilon$ limit. We therefore obtain the one-loop correction to the six-point MHV amplitude of the D3-brane theory, which is given by a contact rational term, 
\begin{align} \label{eq:result}
\mathcal{A}^{\rm MHV}_6 = \,  - \frac{i }{32 \pi^{2}}   \delta^{8}(Q_6) \left( [12]^2[34]^2[56]^2   +{\rm Perm}  \right)  \, .
\end{align}
It is straightforward to see that the above result is the unique answer that has the right power counting and correct little-group scaling, and further require the answer do not possess poles. Indeed,
we do not allow MHV amplitudes to have poles since the theory has no three-point amplitudes. Similarly, at $n$ points, the unique answer that is consistent with the power counting and little-group scaling takes the following form
\begin{align} \label{eq:general}
- \frac{i }{32 \pi^{2}}   \delta^{8}(Q_n) \left( [12]^2[34]^2 \cdots [n{-}1\, n]^2  +{\rm Perm}  \right)  \, . 
\end{align}
The above result is also in agreement with (\ref{eq:npts}) after performing the integral using (\ref{eq:integral}). However, as we commented that the $(n/2)$-gon integral (\ref{eq:npts}) is only a part of the full answer for $n>6$, therefore the overall coefficient of (\ref{eq:general}) has to be determined by explicit computations after including all the lower-topology integrands. 

We would like to comment that the presence of the above rational term for the MHV helicity configuration violates the $U(1)$ symmetry of the tree-level D3-brane amplitudes, which would only allow helicity conserved amplitudes as we commented previously. However, since the rational terms we obtained are purely contact without any poles, they can be simply cancelled by adding local counter terms. \footnote{See similar discussion for the non-supersymmetric BI theory \cite{Elvang:2019twd}.}

Finally, due to the fact that the scalars of D3-brane theory are Goldstone bosons of spontaneously breaking of translation and Lorentz rotation, the corresponding amplitudes with these scalars should obey enhanced soft behaviour  \cite{Cheung:2014dqa,Cheung:2016drk},  
\be
A(k_1, k_2, \cdots, k_n) \big{|}_{k_1 \rightarrow 0} \sim \mathcal{O}(k_1^2)\, , 
\ee
where $k_1$ is the momentum of one of the scalars. The enhanced soft behaviour was further
argued to be valid when the loop corrections are taken into account \cite{Guerrieri:2017ujb}.  To study the soft behaviour of the scalar fields, we consider the rational term with the helicity configuration $(\phi_1, \bar{\phi}_2, 3^-, 4^+, \cdots, n^+)$, where $\phi_1, \bar{\phi}_2$ are scalars. From (\ref{eq:result}), we see that the rational term is then proportional to 
\be
\langle 1 3 \rangle^2 \langle 23 \rangle^2 \left([12]^2[34]^2 \cdots [n{-}1\,n]^2 +\text{Perm}  \right)\, .
\ee
It is easy to see that each term in the permutation goes as $\mathcal{O}(k_1^2)$ (or $\mathcal{O}(k_2^2)$) in the soft limit $k_1 \rightarrow 0$ (or $k_2 \rightarrow 0$), which is consistent with the enhanced soft behaviour of the D3-brane theory. 

\section{Non-supersymmetric D3-brane amplitudes at one loop}
\label{sec:nonsusy}

The loop corrections to scattering amplitudes in 4D lower-supersymmetric theories or non-supersymmetric theories can be obtained by a supersymmetry reduction on 6D tree-level superamplitudes such that only relevant states (instead of the full super multiplets) run in the loops. Using this idea, we will study loop corrections to the amplitudes in the non-supersymmetric Born-Infeld (BI) theory. In particular, for the BI theory, we project the external states to be photons and the internal particles to be a pair of massive vectors.  

In the following sections, we will consider one-loop amplitudes with all-plus external photons (the Self-Dual sector) and single-minus external photons (the Next-to-Self-Dual sector). They vanish identically in the supersymmetric theory, and for the non-supersymmetric theory, they are purely rational terms at one-loop order. Therefore, just as in the case of one-loop MHV amplitudes in the supersymmetric theory, to extract the rational terms we require the internal particles propagating in the loop to be massive. We find that the results from the forward-limit construction agrees with those in \cite{Elvang:2019twd}, which were computed originally using the generalised unitarity methods~\cite{Bern:1994zx, Bern:1994cg}.

\subsection{Self-Dual sector}
We begin with the amplitudes in the Self-Dual (SD) sector, namely the amplitudes with all-plus helicity configuration. We perform a supersymmetric reduction by choosing all external legs to be plus-helicity photons and the forward-limit particles to be a pair of massive vectors or massive scalars\footnote{For the all-plus and single-minus helicity configurations, one can in fact replace the internal massive vectors by massive scalars to simplify the computation, see, for instance, \cite{Elvang:2019twd} for the argument. We have checked that the same results are obtained by choosing the internal states being either massive vectors or massive scalars. }.
With such construction, the one-loop $n$-point amplitude in the SD sector is given by
\be \label{eq:SD4pt}
{A}^{(1)}_{\text{SD},\,n}=\int \frac{d^D\tilde{\ell}}{(2\pi)^D} \frac{1}{{\tilde{\ell}}^2} \int (\prod_{i=1}^{n}d^2\eta_{i,2}) \,d^4 \eta_{n+1}\; \mathcal{A}^{(0)}_{\text{M5},n+2} \big{|}_{\rm F.L.} \,,
\ee
where $d^2\eta_{i,a}=\frac{1}{2}\, \epsilon_{IJ}d\eta_{i,a}^I d\eta_{i,a}^J$. This choice of Grassmann variables integration projects all the external legs (i.e. particles $1$ to $n$) to be plus-helicity photons, and it also sets the internal particles $n{+}1$ and $n{+}2$ to be scalars.

In the case of $n=4$, carrying out the Grassmann integral in (\ref{eq:SD4pt}) and solving the scattering equations, we find that the one-loop correction to the four-point amplitude in the SD sector is given by
\be \label{eq:SD4}
{A}^{(1)}_{\text{SD},\,4}=(\mu^2)^2 [12]^2[34]^2 \,I_{\text{bubble}}(k_1,k_2)+\text{Perm}\, ,
\ee
where the bubble integral $I_{\text{bubble}}(k_1,k_2)$ is given in (\ref{eq:bubble}). 
The above expression agrees with the loop integrand in (4.1) of \cite{Elvang:2019twd}, except for the propagators being linearised. Similar computation applies to higher-point cases. Let us consider the six-point case here, that is given by  (\ref{eq:SD4pt}) with $n=6$, from which we find
\begin{align}  \label{eq:SD6}
{A}^{(1)}_{\text{SD},\,6}=- \int& \frac{d^D\tilde{\ell}}{(2\pi)^D}
(\mu^2)^3 \Big[[12]^2[34]^2[56]^2 \nonumber\\
\times\Big(\; 
&\frac{1}{{\tilde{\ell}}^2\,[2\tilde{\ell} \cdot (k_{1}+k_{2})+2k_1\cdot k_2]\,[-2\tilde{\ell} \cdot (k_5+k_6)+2k_5 \cdot k_6]}\nonumber\\
+&\frac{1}{{\tilde{\ell}}^2\,[2\tilde{\ell} \cdot (k_{3}+k_{4})+2k_3\cdot k_4]\,[-2\tilde{\ell} \cdot (k_1+k_2)+2k_1 \cdot k_2]} \\
+&\frac{1}{{\tilde{\ell}}^2\,[2\tilde{\ell} \cdot (k_{5}+k_{6})+2k_5\cdot k_6]\,[-2\tilde{\ell} \cdot (k_3+k_4)+2k_3 \cdot k_4]}\Big)
+\text{Perm}\Big] \nonumber \,, 
\end{align}
which agrees with the loop integrand in (4.4) of \cite{Elvang:2019twd}. Again our result is in the linear propagator representation, but it is equivalent to that of \cite{Elvang:2019twd} as shown in Appendix \ref{app:llinear}.  

We note that, at least for the cases we studied here, the one-loop integrands of amplitudes in the SD sector of non-supersymmetric D3-brane theory take a very similar form as those of the MHV amplitudes in the supersymmetric D3-brane theory. In fact, they are related to each other by exchanging the factor $(\mu^2)^2$ in amplitudes of the SD sector with $\delta^8(Q)$ in the supersymmetric amplitudes. 

\subsection{Next-to-Self-Dual sector}
The computation for the one-loop amplitudes in the Next-to-Self-Dual (NSD) sector is very similar. They are the amplitudes with single-minus helicity configuration, so we only need to change the choice of Grassmann variables from all-plus to single-minus. Applying a similar construction as (\ref{eq:SD4pt}), we have
\be
{A}^{(1)}_{\text{NSD},\,n}=\int \frac{d^D\tilde{\ell}}{(2\pi)^D} \frac{1}{{\tilde{\ell}}^2}
\int d^2\eta_{n,1}(\prod_{i=1}^{n-1}d^2\eta_{i,2}) \,d^4 \eta_{n+1}\;\mathcal{A}^{(0)}_{\text{M5}, n+2} \big{|}_{\rm F.L. } \,.
\ee 
We assign particle-$n$ to be a minus-helicity photon and the rest of the external particles to be plus-helicity photons. The forward-limit particles $n{+}1$ and $n{+}2$ are again chosen to be scalars.

Let us consider explicitly the four- and six-point amplitudes in the NSD sector. In the case of $n=4$, we find that the result has the form
\begin{align} \label{eq:inte1}
& {A}^{(1)}_{\text{NSD},\,4}\nonumber\\[0.2cm]
&=\int \frac{d^D\tilde{\ell}}{(2\pi)^D} \left( \frac{\mu^2\,[12]^2\,\langle 4|\tilde{\ell} |3]^2}{{\tilde{\ell}}^2[2\tilde{\ell} \cdot (k_1+k_2)+2k_1\cdot k_2]}+\frac{\mu^2\,[12]^2\,\langle 4| \tilde{\ell}-(k_1+k_2)|3]^2}{{\tilde{\ell}}^2[-2\tilde{\ell} \cdot (k_1+k_2)+2k_1\cdot k_2]}
\right)+\text{Perm}\, ,
\end{align}
where we have defined $\langle i |k  | j ] = \lambda_i^{\a}\,  k_{\a \dot \a} \, \tilde{ \lambda}_{j}^{\dot \a} \, .$ The result is in agreement with the one in (4.10) of \cite{Elvang:2019twd},  as shown in Appendix \ref{app:llinear}. For the six-point amplitude, we find the contributions contain triangle and bubble diagrams, 
\be
{A}^{(1)}_{\text{NSD},\,6}=\int \frac{d^D\tilde{\ell}}{(2\pi)^D}\Big(\mathcal{I}^{\text{triangle}}_{\text{NSD},6}+\mathcal{I}^{\text{bubble}}_{\text{NSD},6}\Big) \,.
\ee
The triangle contribution is given by
\begin{align} \label{eq:inte2}
\mathcal{I}^{\text{triangle}}_{\text{NSD},6}=- (\mu^2)^2 \, &\Big[ \; \frac{[12]^2[34]^2\langle 6| \tilde{\ell} |5]^2}{{\tilde{\ell}}^2[2\tilde{\ell} \cdot (k_{1}+k_{2})+2k_1\cdot k_2]\,[-2\tilde{\ell} \cdot (k_5+k_6)+2k_5 \cdot k_6]}\nonumber\\
&+\frac{[12]^2[34]^2\langle 6| \tilde{\ell}-(k_1+k_2)|5]^2}{{\tilde{\ell}}^2[2\tilde{\ell} \cdot (k_{3}+k_{4})+2k_3\cdot k_4]\,[-2\tilde{\ell} \cdot (k_1+k_2)+2k_1 \cdot k_2]} \\
&+\frac{[12]^2[34]^2\langle 6| \tilde{\ell}+(k_5+k_6)|5]^2}{{\tilde{\ell}}^2[2\tilde{\ell} \cdot (k_{5}+k_{6})+2k_5\cdot k_6]\,[-2\tilde{\ell} \cdot (k_3+k_4)+2k_3 \cdot k_4]} +\text{Perm}  \Big]  \,, \nonumber
\end{align}
and the bubble integral takes the following form
\begin{align}
\mathcal{I}^{\text{bubble}}_{\text{NSD},6}=- (\mu^2)^2\,
 \Big[ & \frac{[12]^2[34]^2\,\langle 6|(k_1+k_2)|5]^2}{s_{125}\,{\tilde{\ell}}^2\,[2\tilde{\ell} \cdot (k_1+k_2)+2k_1\cdot k_2]}
 \nonumber \\[0.2cm]
 +&\frac{[12]^2[34]^2\,\langle 6|(k_1+k_2)|5]^2}{s_{125}\,{\tilde{\ell}}^2\, [-2\tilde{\ell} \cdot (k_1+k_2)+2k_1\cdot k_2]}
+\text{Perm} \Big] \,.
\end{align}
The results are in the agreement with (4.17) and (4.19) of \cite{Elvang:2019twd}, after translating the quadratic propagators into the linear ones as discussed in Appendix \ref{app:llinear}.  

Finally, we comment that the above supersymmetric reduction procedure is very general, and it can be applied to other non-supersymmetric theories. For instance, we have checked explicitly that the procedure reproduces well-known results of some rational terms in pure Yang-Mills theory~\cite{Bern:1995db}. They are obtained from the supersymmetric reduction of the amplitude in 6D SYM with the forward limit as we described.

\section*{Acknowledgements}

We would like to thank Ricardo Monteiro and Ricardo Stark-Much\~ ao for helpful discussions. C.W. is supported by a Royal Society University Research Fellowship No. UF160350.  S.Q.Z. is supported by the Royal Society grant RGF\textbackslash R1\textbackslash 180037.
\appendix

\section{Linear and quadratic  propagators}
\label{app:llinear}

In this appendix, we review the equivalence of the linear-propagator and quadratic-propagator of the $m$-gon integral following mostly \cite{Huang:2015cwh}. The main idea is that we shift the loop momentum in the quadratic $m$-gon as the following
\begin{align} \label{eq:ellshift}
\tilde{\ell} \rightarrow \tilde{\ell}+\beta \,,
\end{align}
where $\beta$ is in extra dimension other than $\tell$; that is, $\tell\cdot\beta=0$ and $k_i \cdot \beta=0$ for an external momentum $k_i$. So the propagator is deformed as
\begin{align} \label{eq:ellshift}
\tilde{\ell}^2\rightarrow(\tilde{\ell}+\beta)^2= \tell^2+\beta^2= \tell^2+z \,,
\end{align}
where we define $z:=\beta^2$. With such shift, we can write a loop integral as a Cauchy integral surrounding the $z=0$ pole in the same spirit of BCFW recursion relation~\cite{Britto:2005fq}
\be
\mathcal{I}(\tell) = \oint_{z=0} \frac{dz}{z}\, \mathcal{I}(\hat{\ell}) \,,
\ee
where $\mathcal{I}(\tell)$ is the one-loop integrand in quadratic form, and shifted loop momentum $\hat{\ell}$ is defined as $\hat{\ell}=\tell+\beta$.
We then apply Cauchy's theorem to deform the contour and pick up (minus) the residues except for that at the pole of $z=0$, 
\be
\oint_{z=0} \frac{dz}{z}\, \mathcal{I}(\hat{\ell})=(-1)\sum \text{Res}\big[\,\frac{\mathcal{I}(\hat{\ell})}{z}\,\big] \, ,
\ee
Each of the residues contains the following factor
\be
\frac{1}{(\tell+K_0)^2},
\ee
with $K_0$ being some 4D momentum. Since we need to perform loop integration $\int\frac{d^D\tell}{(2\pi)^D}$ at the end, we are free to shift the loop momentum to obtain another integral which is equivalent to the original one upon loop integration. So we can do the following change of variable
\be
\tell\rightarrow \tell-K_0\,,
\ee
which sets $\frac{1}{(\tell+K_0)^2}\rightarrow\frac{1}{\tell^2}$. Recall $\tell = \ell + \mu$, where $\ell$ is the 4D loop momentum, therefore the above shift only affects  $ \ell$ but not $\mu$.  The result is an expression that only $\frac{1}{\tell^2}$ is quadratic, while the other propagators are linearised. 

Let us use the bubble diagram as an explicit example to illustrate the above procedure, in the quadratic form it is given by
\begin{align} 
\tilde{I}_{\text{bubble}}(k_{1},k_{2})=\int \frac{d^{D} \tilde{\ell} }{(2\pi)^{D}}\, &\Big[ \;\frac{\mu^{m}}{\tilde{\ell}^2 \, (\tilde{\ell} + k_{1}+k_{2} )^2} \Big] \, .
\end{align}
Using the shift (\ref{eq:ellshift}) and dividing by $z$, the above bubble integral can be expressed as an contour integral surrounding $z=0$ pole 
\begin{align} 
\tilde{I}_{\text{bubble}}(k_{1},k_{2})= \int \frac{d^{D} \tilde{\ell} }{(2\pi)^{D}}\, \oint_{z=0} dz \Big[ \;\frac{\mu^{m}}{z\,(\tilde{\ell}^2+z) \, ({\tilde{\ell}}^2+z + 2\tell\cdot k_{1,2} +k_{1,2}^2)} \Big] \, , 
\end{align}
where $k_{i,j}= k_i + k_j$. Applying Cauchy's theorem, we find that two residues contribute as
\begin{align}
\int \frac{d^{D} \tell}{(2\pi)^{D}}\,  \Big[ \frac{\mu^{m}}{ \tell^2 (2\tell\cdot k_{1,2} +k_{1,2}^2)} +\frac{\mu^{m}}{ (\tell+k_{12})^2 (-2\tell\cdot k_{1,2} -k_{1,2}^2)}\Big] \, .
\end{align}
The shift of the loop momentum in the second term results into the linear-propagator representation of the bubble integral
\begin{align}
\tilde{I}_{\text{bubble}}(k_{1},k_{2})\simeq \int \frac{d^{D} \tell}{(2\pi)^{D}}\,  \Big[ \frac{\mu^{m}}{ \tell^2 (2\tell\cdot k_{1,2} +k_{1,2}^2)} +\frac{\mu^{m}}{ \tell^2 (-2\tell\cdot k_{1,2} +k_{1,2}^2)}\Big] \, ,
\end{align}
where $\simeq$ denotes equivalence upon integration.  

The linearised triangle can be obtained by similar manners, we first write the quadratic one as an Cauchy integral
\be 
\oint_{z=0}dz\int \frac{d^D\tilde{\ell}}{(2\pi)^D}\,
\left[ \frac{\mu^m}{z\,(\tell^2+z)(\tell^2+z+2\tell\cdot k_{1,2}+k_{1,2}^2)\,(\tell^2+z-2\tell \cdot k_{5,6}+k_{5,6}^2)} \right] \, .
\ee
We again apply the Cauchy theorem and shift loop momentum to obtain the following expression
\begin{align}  
\int \frac{d^D\tilde{\ell}}{(2\pi)^D}\;
\Big[\; 
&\frac{\mu^{m}}{{\tilde{\ell}}^2\,(2\tilde{\ell} \cdot k_{1,2}+k_{1,2}^2)\,(-2\tilde{\ell} \cdot k_{5,6}+k_{5,6}^2)}\nonumber\\
+&\frac{\mu^{m}}{{\tilde{\ell}}^2\,(2\tilde{\ell} \cdot k_{3,4}+k_{3,4}^2)\,(-2\tilde{\ell} \cdot k_{1,2}+k_{1,2}^2)} \\
+&\frac{\mu^{m}}{{\tilde{\ell}}^2\,(2\tilde{\ell} \cdot k_{5,6}+k_{5,6}^2)\,(-2\tilde{\ell} \cdot k_{3,4}+k_{3,4}^2)}\,\Big] \, . \nonumber
\end{align}

Similar computation applies to the integrals with non-trivial numerators such as those given in (\ref{eq:inte1}) and (\ref{eq:inte2}). In particular, one can show that (\ref{eq:inte1}) is equivalent to 
\begin{align} 
\tilde{\mathcal{A}}^{(1)}_{\text{NSD},\,4} =\int \frac{d^D\tilde{\ell}}{(2\pi)^D}  \frac{\mu^2\,[12]^2\,\langle 4|\tilde{\ell} |3]^2}{{\tilde{\ell}}^2 (\tilde{\ell} + k_1+k_2)^2}+\text{Perm}\, ,
\end{align} 
and under the loop integration, (\ref{eq:inte2}) is equivalent to 
\begin{align} 
\tilde{\mathcal{I}}^{\text{triangle}}_{\text{NSD},6}=   \frac{ (\mu^2)^2 [12]^2[34]^2\langle 6| \tilde{\ell} |5]^2}{{\tilde{\ell}}^2\, (\tilde{\ell} + k_{1}+k_{2})^2 \, (\tilde{\ell} - k_5 - k_6)^2} +\text{Perm}  \, . 
\end{align}



\end{document}